\documentclass[aps,pre,showpacs,floatfix,twocolumn,10pt]{revtex4-1}

\usepackage{amssymb,amsfonts,amsmath}
\usepackage{bm}				
\usepackage{latexsym}
\usepackage{graphicx}

\usepackage{color}

\definecolor{myblue}{RGB}{65,105,225}
\definecolor{mygreen}{RGB}{34,139,34}
\definecolor{myorange}{RGB}{255,69,0}
\usepackage[colorlinks,linkcolor=myorange,urlcolor=myblue,citecolor=myblue]{hyperref}

\usepackage[table,xcdraw]{xcolor}

\def\rr{\textcolor{black}}

\newcommand{\be}{\begin{equation}}
\newcommand{\ee}{\end{equation}}
\newcommand{\la}{\langle}
\newcommand{\ra}{\rangle}
\newcommand{\ben}{\begin{eqnarray}}
\newcommand{\een}{\end{eqnarray}}

\newcommand{\talpha}{\gamma}
\newcommand{\tdelta}{\delta}

\def\(({\left(}
\def\)){\right)}
\def\[[{\left[}
\def\]]{\right]}

\begin{document}

\title{A violation of universality in anomalous Fourier's law}

\author{Pablo I. Hurtado}
\email{phurtado@onsager.ugr.es}
\affiliation{Institute Carlos I for Theoretical and Computational Physics and Departamento de Electromagnetismo y F\'isica de la Materia, Universidad de Granada, 18071 Granada, Spain}
\author{Pedro L. Garrido}
\email{garrido@onsager.ugr.es}
\affiliation{Institute Carlos I for Theoretical and Computational Physics and Departamento de Electromagnetismo y F\'isica de la Materia, Universidad de Granada, 18071 Granada, Spain}

\begin{abstract} 
Since the discovery of long-time tails, it has been clear that Fourier's law in low dimensions is typically anomalous, with a size-dependent heat conductivity, though the nature of the anomaly remains puzzling. The conventional wisdom, \rr{supported by renormalization-group arguments and mode-coupling approximations within fluctuating hydrodynamics}, is that the anomaly is universal in $1d$ momentum-conserving systems and belongs in the \rr{L\'evy/Kardar-Parisi-Zhang} universality class. Here we challenge this picture by using a novel scaling method to show unambiguously that universality breaks down in the paradigmatic $1d$ diatomic hard-point fluid. Hydrodynamic profiles for a broad set of gradients, densities and sizes all collapse onto an universal master curve, showing that (anomalous) Fourier's law holds even deep into the nonlinear regime. This allows to solve the macroscopic transport problem for this model, a solution which compares flawlessly with data and, interestingly, implies the existence of a bound on the heat current in terms of pressure. \rr{These results question the renormalization-group and mode-coupling universality predictions for anomalous Fourier's law in $1d$, offering a new perspective on transport in low dimensions.}
\end{abstract}

\date{\today}


\maketitle

\section{Introduction}

It's going to be 200 years since Fourier stated his seminal law \cite{fourier1822a}, but its microscopic understanding still poses one of the most important and challenging open problems in nonequilibrium statistical physics, with no rigorous mathematical derivation to date \cite{lebowitz11a,bonetto00a,lepri03a,dhar08a,lepri16a,liu12a}. Fourier's law establishes the proportionality between the heat current and the local temperature gradient in a material, with the proportionality factor defining the heat conductivity $\kappa$, a key material property. While for bulk, three-dimensional materials $\kappa$ is well characterized and measured, its status in low-dimensional structures is far from clear. In particular, for low-dimensional systems ($d=1,2$) with momentum conservation, \rr{the effective conductivity} $\kappa$ grows with the system size $L$, diverging in the thermodynamic limit and thus leading to anomalous heat transport \cite{bonetto00a,lepri03a,dhar08a,liu12a,lepri16a}. The understanding of this anomaly has attracted a lot of attention in recent years \cite{bonetto00a,lepri03a,dhar08a,liu12a,lepri16a,ghosh08a,balandin08a,ghosh10a,balandin11a,xu14a,meier14a,chang08a,henry08a,liu12b,hsiao13a,yang10a,huang12a,alder67a,alder70a,resibois77a,narayan02a,beijeren12a,mendl13a,spohn14a,das14a,mendl15a,popkov15a,lee-dadswell15a,delfini06a,delfini07a,delfini07b,politi11a,pozo15b,hurtado06a,hurtado05a,lee-dadswell05a,lee-dadswell15b,liu14a,li15a}, not only because it is expected to shed light on the key ingredients behind Fourier's law at a fundamental level, but also because of its technological relevance in low-dimensional real-world materials, the most noteworthy being graphene \cite{ghosh08a, balandin08a, ghosh10a, balandin11a,xu14a}, but with other important examples ranging from molecular chains \cite{meier14a} and carbon nanotubes \cite{chang08a} to polymer fibers \cite{henry08a,liu12b}, nanowires \cite{hsiao13a,yang10a} and even spider silk \cite{huang12a}, to mention just a few; see \cite{liu12a} for a recent review. From a theoretical perspective, the low-dimensional anomaly in heat transport can be linked to the presence of strong dynamic correlations in these fluids and lattices \cite{alder67a,alder70a,resibois77a}, though a detailed understanding has remained elusive for decades. 

\rr{
In $1d$, clear signatures of this anomaly appear in a number of different phenomena \cite{lepri16b}. For instance, the steady state heat current $J$ of a $1d$ momentum-conserving system driven by a \emph{small} boundary temperature gradient (i.e. in the linear regime) typically scales as $L^{-1+\talpha}$ for large enough system sizes $L$, with $0\le\talpha<1$ an anomaly exponent, which can be interpreted in terms of a finite-size heat conductivity $\kappa_L \sim L^\talpha$. An exponent $\talpha=0$ corresponds to standard diffusive transport, but typically $\talpha>0$ is observed in $1d$ implying superdiffusive heat transport \cite{lepri16b}. The low-dimensional transport anomaly is also apparent in equilibrium. In particular, the long-time tail of the equilibrium time correlation of the energy current decays in $1d$ in a nonintegrable, power-law way, $\la J(0) J(t)\ra \sim t^{-1+\tdelta}$ as $t\to\infty$, with $0\le\tdelta<1$ another exponent. Green-Kubo relations for the transport coefficients hence imply a divergent value for the heat conductivity, in agreement with nonequilibrium results \cite{lepri16a}. Additional signatures of anomalous transport have been also reported in the superdiffusive spreading of energy perturbations in equilibrium \cite{cipriani05a,lepri16b,liu14a,li15a}. A range of different values for the exponents $\talpha$ and $\tdelta$ have been measured in simulations and experiments for different model systems \cite{lepri03a,dhar08a,liu12a,lepri16a}, the main difficulty being extracting the large $L$ asymptotics due to the strong and poorly understood finite-size effects affecting these measurements (which mix bulk and boundary finite-size corrections). The prevailing picture, however, is that the transport anomaly exponents are \emph{universal} and within the L\'evy/Kardar-Parisi-Zhang (L/KPZ) universality class \cite{bonetto00a,lepri03a,dhar08a,liu12a,lepri16a}, a conjecture based on renormalization-group \cite{narayan02a} and mode-coupling \cite{beijeren12a} calculations, and reinforced by recent related breakthroughs from nonlinear fluctuating hydrodynamics \cite{mendl13a,spohn14a,das14a,mendl15a,popkov15a} which predict L\'evy (KPZ) scaling for the heat (sound) modes of the equilibrium time correlators of \emph{conserved fields}. In particular, for the transport anomaly $\talpha=1/3=\tdelta$ is expected in the general case, though a second universality class with $\talpha=1/2=\tdelta$ \rr{seems to} appear under special circumstances (as e.g. for zero-pressure systems with symmetric potential \cite{beijeren12a,lee-dadswell15a,delfini06a,delfini07a,delfini07b,politi11a}). Special cases with convergent $\kappa$ in $1d$, as the coupled rotors model \cite{giardina00a,gendelman00a}, can be also accounted for by fluctuating hydrodynamics after noticing that these models have less than three locally-conserved fields \cite{das14b}.
}

In this work we challenge the \rr{universality conjecture for anomalous Fourier's law} by using a novel scaling method to offer a high-precision measurement of the conductivity anomaly in a paradigmatic $1d$ model of transport. Compared to previous attempts at characterizing the \rr{transport} anomaly, most based on \emph{linear} response theory and hence critically-dependent on a large system-size limit (which is in fact never attained) \cite{lee-dadswell15b}, our method takes full advantage of the \emph{nonlinear} character of the heat conduction problem in a natural way, \rr{allowing to disentangle the crucial bulk size scaling from the artificial boundary finite-size corrections}. Our model is the archetypical $1d$ diatomic hard-point gas in a temperature gradient \cite{casati86a, garrido89a, garrido01a, savin02a, grassberger02b, casati09a, brunet10a, boozer11a, mendl14a, chen14a,chen14b}, which is characterized by the mass ratio $\mu=M/m>1$ between neighboring particles. We unambiguously show below that, contrary to the standard lore, this model does obey \rr{an anomalous version of} Fourier's law, namely 
\be
J=-\kappa_L(\rho,T)\frac{dT(x)}{dx} \, ,
\label{fourierL}
\ee
for a broad range of temperature gradients (from the linear response domain to the deeply nonlinear regime), with the heat current $J$ proportional to the local temperature gradient via a conductivity functional
\be
\kappa_L(\rho,T)=L^\alpha \sqrt{\frac{T}{m}} k(\rho) \, .
\label{conductL}
\ee
\rr{Note that Eqs. (\ref{fourierL})-(\ref{conductL}) are not Fourier's law in the usual sense, as the latter implies a size-independent $\kappa$, while the conductivity in this case grows with the system size as $L^\alpha$, with $\alpha$ a new exponent characterizing anomalous transport in $1d$. The validity of Eqs. (\ref{fourierL})-(\ref{conductL})} is proven \rr{below} by collapsing onto a striking universal master curve the density and temperature profiles measured for a large set of system sizes, number densities and temperature gradients. Such compelling collapse offers a high-precision measurement of the anomaly exponent $\alpha$, which remarkably turns out to be \emph{non-universal}, depending non-monotonously on the mass ratio $\mu$. The observed scaling allows to solve the macroscopic transport problem for this model, and we obtain analytic expressions for the universal master curve (\rr{as well as for} the hydrodynamic profiles, \rr{current, pressure, etc.}) which exhibit an excellent agreement with measurements. Interestingly, this solution immediately implies the existence of a nontrivial bound on the current in terms of pressure $P$. 

A natural question concerns the relation of the new anomaly exponent $\alpha$ with the standard ones defined in literature, namely $\gamma$ in the linear response regime and $\delta$ from equilibrium current time correlations (see description above). This relation can be easily established by studying the linear response limit of the anomalous Fourier's law (\ref{fourierL})-(\ref{conductL}), a particular regime of the broad range of temperature gradients where these equations hold with high accuracy, as we demonstrate below. Indeed, for small enough boundary temperature difference $\Delta T$ the local temperature gradient can be written as $dT/dx\approx -\Delta T/L$, and this together with Eqs. (\ref{fourierL})-(\ref{conductL}) leads to $J\propto L^{-1+\alpha}$, an argument which strongly suggests the conjecture $\alpha=\gamma(=\delta)$. In this way, the surprising but clear-cut dependence of $\alpha$ on the mass ratio $\mu$ reported below hence signals the breakdown of the universality claimed for $1d$ anomalous Fourier's law. We maintain here however the different notation for the various (but related) anomaly exponents to stress out their distinct definitions.

\section{Results}

\begin{figure}
\vspace{-0.3cm}
\centerline{\includegraphics[width=9cm]{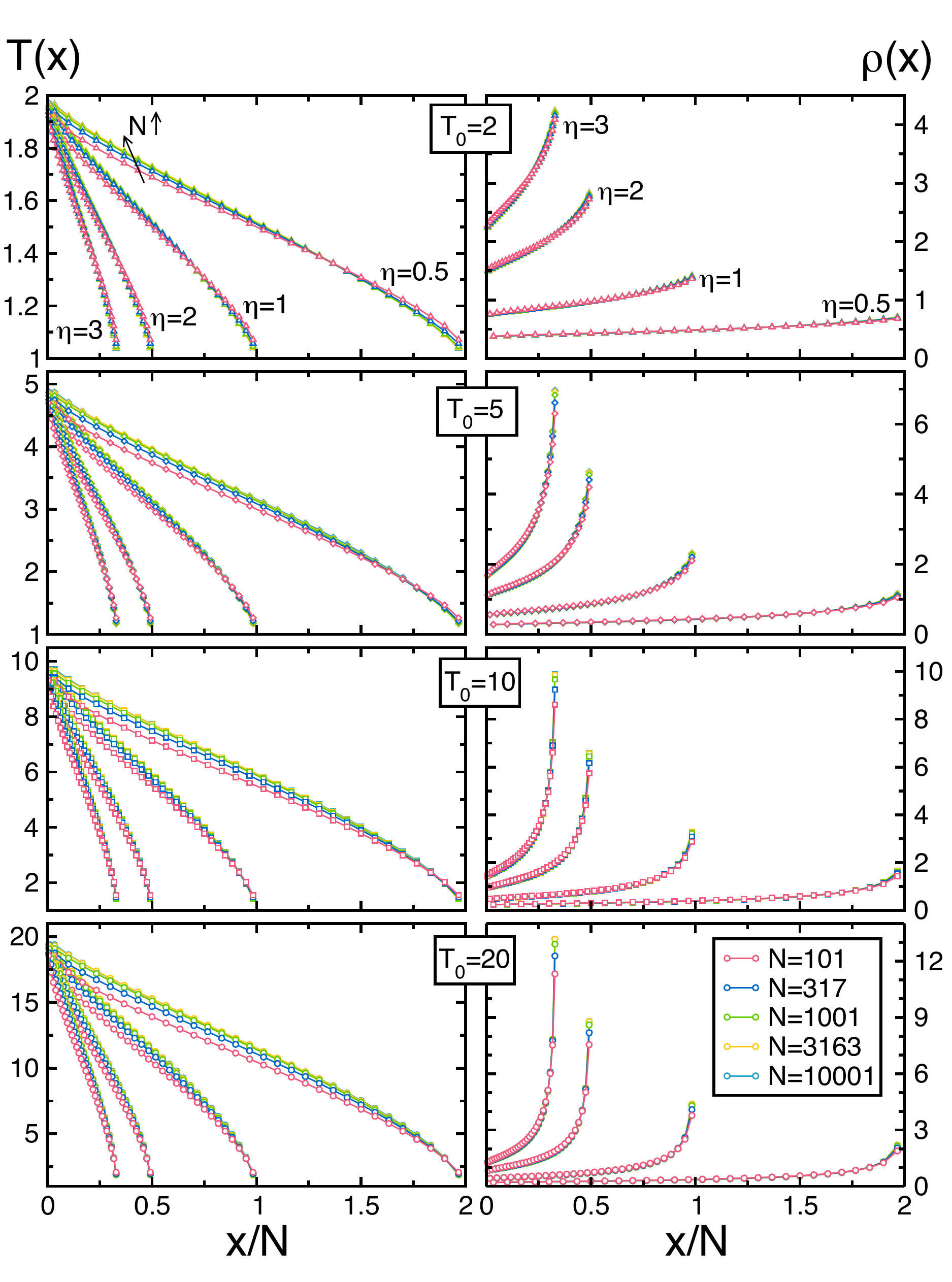}}
\vspace{-0.3cm}
\caption{\small  {\bf Temperature and density fields under a thermal gradient.} Temperature (left) and density (right) profiles measured for (from top to bottom) $T_0=2, 5, 10, 20$ and varying $\eta$ and $N$, for a mass ratio $\mu=3$. 
}
\label{fig1}
\end{figure}

\begin{figure*}[t]
\vspace{-0.3cm}
\centerline{\includegraphics[width=18.cm]{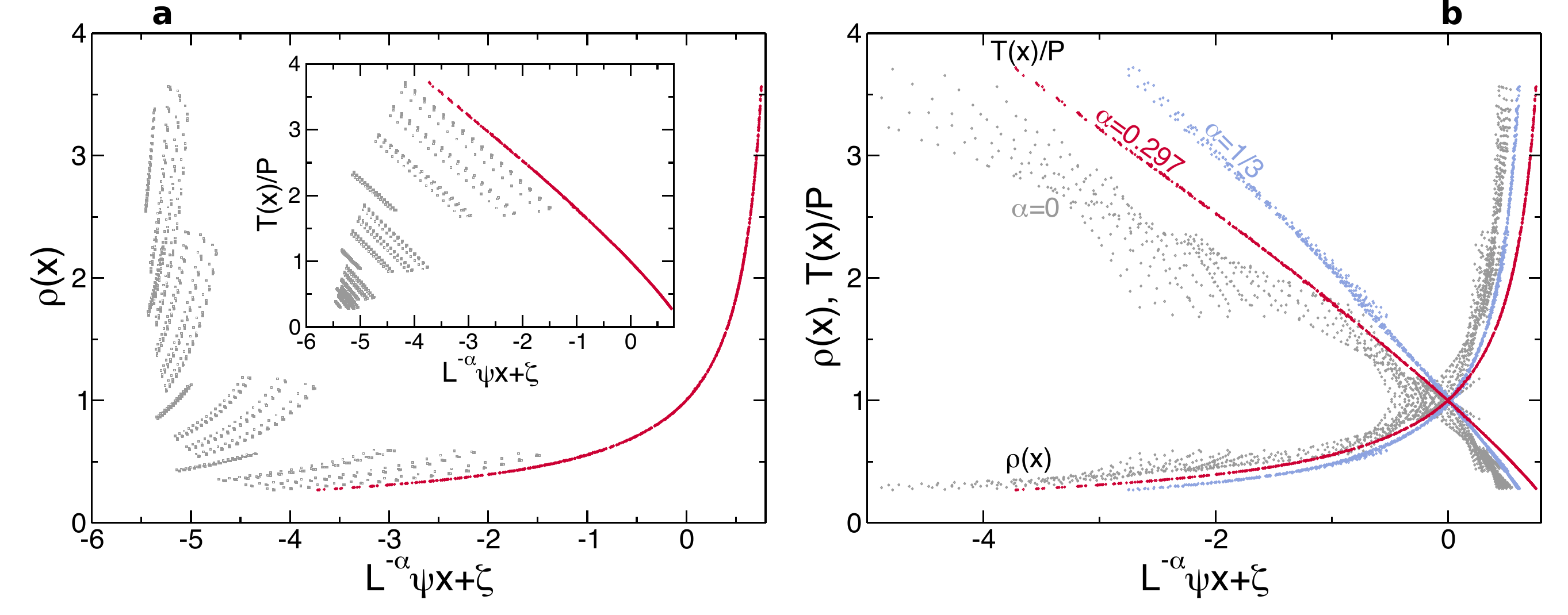}}
\vspace{-0.3cm}
\caption{\small  {\bf Scaling procedure and data collapse.}
(a) Density profiles for $\mu=3$ $\forall N, T_0, \eta$ as a function of $L^{-\alpha}\psi x$ with $\alpha=0.297$, before (light gray) and after (dark red) the shifts $\zeta$. Inset: Same as before, but for the reduced temperature profiles. Note that the shifts are those obtained from density profiles. In both cases the data collapse is remarkable. 
(b) Optimal collapse of density and reduced temperature profiles for $\mu=3$ and three different exponents $\alpha=0, 0.297$, and $1/3$. The superior collapse for $\alpha=0.297$ is apparent. The abscisa for $\alpha=0$ has been divided by a factor 10 for the sake of clarity.
}
\label{fig2}
\end{figure*}

We hence consider a $1d$ Hamiltonian model fluid consisting in $N$ hard-point particles with alternating masses, $m=1$ and $M=\mu m>1$, moving ballistically in a line of length $L$ in between elastic collisions with neighboring particles. The fluid is coupled to two stochastic thermal walls at the boundaries, $x=0,L$, which reflect particles upon collision with a velocity modulus randomly drawn from a Maxwellian distribution defined by the wall temperature $T_{0,L}$  \cite{bonetto00a,lepri03a,dhar08a,lepri16a}. \rr{For $T_0\ne T_L$,} the temperature gradient drives the system to an inhomogeneous nonequilibrium steady state characterized by nonlinear density and temperature profiles, $\rho(x)$ and $T(x)$ respectively \cite{bonetto00a,lepri03a,dhar08a,liu12a,lepri16a}. Interestingly, these profiles can be shown to follow from an universal master curve, independent of the driving gradient and the fluid's density, if and only if (i) Fourier's law (\ref{fourierL})\rr{-(\ref{conductL})} and (ii) macroscopic local equilibrium (MLE) hold (see Section I of the Supplementary information for a detailed proof), an equivalence which holds for general $d$-dimensional systems \cite{pozo15b}. MLE implies that the stationary density and temperature fields are locally coupled via the equilibrium equation of state (EoS) \cite{pozo15a}, which for the $1d$ diatomic hard-point fluid simply takes the ideal gas form, $P=\rho T$. In this way, iff hypotheses (i)-(ii) hold, we expect all density and temperature profiles to scale as
\be
\rho(x)=F\left(\frac{\psi x}{L^{\alpha}} +\zeta\right) \quad ; \quad \frac{T(x)}{P}=1/F\left(\frac{\psi x}{L^{\alpha}} +\zeta\right) \,
\label{scale}
\ee
with 
$\psi=J\sqrt{m}/P^{3/2}$ the reduced current and $\zeta$ a constant, see Section I of the Supplementary information. This scaling defines an universal master curve $F(u)$ from which all profiles follow. Alternatively, Eq. (\ref{scale}) implies that all measured density and temperature profiles can be collapsed onto an universal master curve after appropriately scaling space by $L^{-\alpha}\psi$, with $\psi$ measured in each case, and shifting the curve by a constant $\zeta$. The resulting collapse is expected to be very sensitive to the anomaly exponent $\alpha$, and this suggests a simple scaling procedure to measure both $\alpha$ and the universal master curve in simulations, confirming at the same time our starting hypotheses.

In order to do so, we performed a large number of event-driven simulations of the $1d$ diatomic gas for a broad set of boundary temperatures $T_0=2,5,10,20$ (with fixed $T_L=1$), global number densities $\eta\equiv N/L=0.5, 1, 2, 3$, different mass ratios $\mu=1.3, 1.618, 2.2, 3, 5, 10, 30, 100$, and numbers of particles $N=101, 317, 1001, 3163, 10001$, reaching up to $N=10^5+1$ in some cases. We measured locally a number of relevant observables including the local kinetic energy, number density, virial pressure and energy current density, as well as the energy current flowing through the thermal reservoirs at $x=0,L$ and the pressure exerted on these walls. We stress that observables measured at the walls agree in all cases with their bulk counterparts, which are constant along the system. For local measurements, we divided the fluid in 30 virtual cells, a constant number independent of other system parameters. The simulation time unit was set to $t_0=\sqrt{M/(2T_L\eta^2)}$, the mean free time of a heavy particle in a cool environment, and time averages were performed taking into account the relaxation and correlation timescales of the $1d$ fluid, which grow strongly with $N$ (see \rr{Fig. S4 and Section II.B in the Supplementary information}). Statistical errors are computed in all cases at $99.7\%$ confidence level, and error bars are shown if larger than the plotted symbols. 

\begin{figure*}[t]
\vspace{-0.3cm}
\centerline{
\includegraphics[width=18.cm]{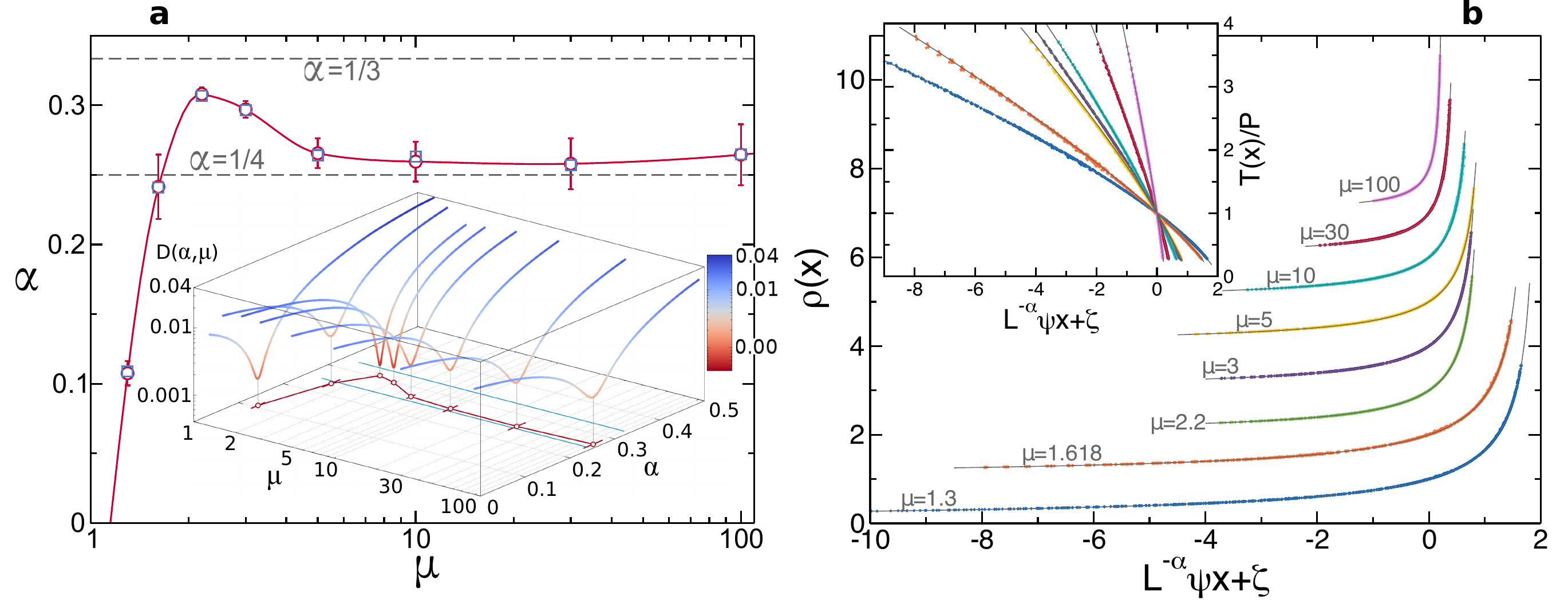}
}
\vspace{-0.3cm}
\caption{\small  {\bf Breakdown of universality and master curves in anomalous Fourier's law.}
(a) Mass ratio dependence of the anomaly exponent measured from scaling ($\bigcirc$). The non-monotonous behavior of $\alpha(\mu)$ clearly signals the breakdown of universality for anomalous Fourier's law in $1d$. The exponent measured from the power-law fit for $k(\rho)$ is also shown ($\Box$), being fully compatible with $\alpha$ in each case. The line is a guide to the eye. Inset: The collapse metric $D(\alpha,\mu)$ as a function of $\alpha$ exhibits a deep and narrow minimum for each $\mu$ (note the logarithmic scale in $z$-axis), offering a precise measurement of the anomaly exponent and its error.
(b) Collapse of density profiles for each $\mu$ obtained by using the measured $\alpha$ in each case. The master curves have been shifted vertically for better comparison. In all cases, the data collapse is excellent. The lines are theoretical predictions, see main text. Inset: Collapse of reduced temperature profiles for the same conditions, and theoretical curves. In all cases, each curve for fixed $\mu$ contains 1280 points measured in 80 different simulations for varying $N$, $T_0$ and $\eta$. The abscisas for the $\mu=1.3$ data have been divided by 4 to better visualize the results. 
}
\label{fig3}
\end{figure*}

Fig. \ref{fig1} shows the temperature and density profiles measured for $\mu=3$ and varying $T_0$, $\eta$ and $N$ (similar data are obtained for all other $\mu$'s). These profiles are clearly nonlinear, and exhibit strong finite-size effects. However, the measured local density and temperature in each case are tightly coupled by the equilibrium EoS, $P=\rho(x)T(x)$, with $P$ the finite-size pressure measured in each simulation, \rr{see Fig. S2 and Section II.A in the Supplementary information}, validating hypothesis (ii) above and confirming the robustness of MLE far from equilibrium \cite{pozo15a}. Note that the thermal walls act as defects (akin to fixed, infinite-mass particles) which disrupt the structure of the surrounding fluid, defining two boundary layers where finite-size corrections mount up. To analyze below the fluid's scaling behavior, we neglect data from these boundary layers (up to 7 cells adjacent to each wall), focusing the analysis on the remaining \emph{bulk} profiles $\rho(x)$ and $T(x)$. For a given exponent $\alpha$, each bulk density profile $\rho(x)$ is then plotted as a function of $L^{-\alpha}\psi x$ (with $\psi=J\sqrt{m}/P^{3/2}$ measured in each case, see \rr{Supplementary Fig. S3}), and shifted by a constant $\zeta$ to achieve an optimal collapse among all scaled profiles, see Fig. \ref{fig2}.a. The vector of optimal shifts $\bm{\zeta}_0$ for fixed $\alpha$ and $\mu$ is obtained by minimizing a standard collapse metric $D(\bm{\zeta};\alpha,\mu)$ for the density profiles \rr{(defined in detail in Section III of the Supplementary information)}, which roughly speaking measures the relative average distance among all pairs of overlapping curves \cite{bhattacharjee01a}, and the same shifts are used to collapse reduced temperature profiles, $T(x)/P$. The resulting data collapses are very sensitive to $\alpha$, \rr{see Fig. \ref{fig2}.b}, so the the true anomaly exponent $\alpha$ can be measured with high precision for each mass ratio $\mu$ by minimizing $D(\alpha,\mu)\equiv D(\bm{\zeta}_0;\alpha,\mu)$ as a function of $\alpha$. In fact, the distance function $D(\alpha,\mu)$ has a pronounced minimum in $\alpha$ for each $\mu$, see inset in Fig. \ref{fig3}.a, whose width and depth allow to estimate the exponent error, see \rr{Supplemementary information, Section III}. 

Remarkably, the measured anomaly exponent is \emph{non-universal}, depending non-monotonously on the mass ratio, $\alpha=\alpha(\mu)$, see Fig. \ref{fig3}.a \rr{and Supplementary Table S1}, growing first from small values at low $\mu$ to a maximum $\alpha\approx 0.3<1/3$ for $\mu=2.2$, and decaying afterwards to an asymptotic value $\alpha\approx 1/4$ for large $\mu$. Fig. \ref{fig3}.b shows the master curves obtained from density and reduced temperature bulk profiles for different $\mu$'s by using the measured $\alpha$'s, and in all cases the resulting collapses are impressive, confirming that \rr{anomalous} Fourier's law (\ref{fourierL})\rr{-(\ref{conductL})} rules heat transport in this $1d$ model. Moreover, this surprising but unambiguous result also calls into question the prevailing \rr{conjecture} that the anomaly in $1d$ heat transport is universal \cite{narayan02a,beijeren12a,mendl13a,spohn14a,das14a,mendl15a,popkov15a,lee-dadswell15a,lepri16a}.

\begin{figure*}
\vspace{-0.3cm}
\centerline{
\includegraphics[width=18.cm]{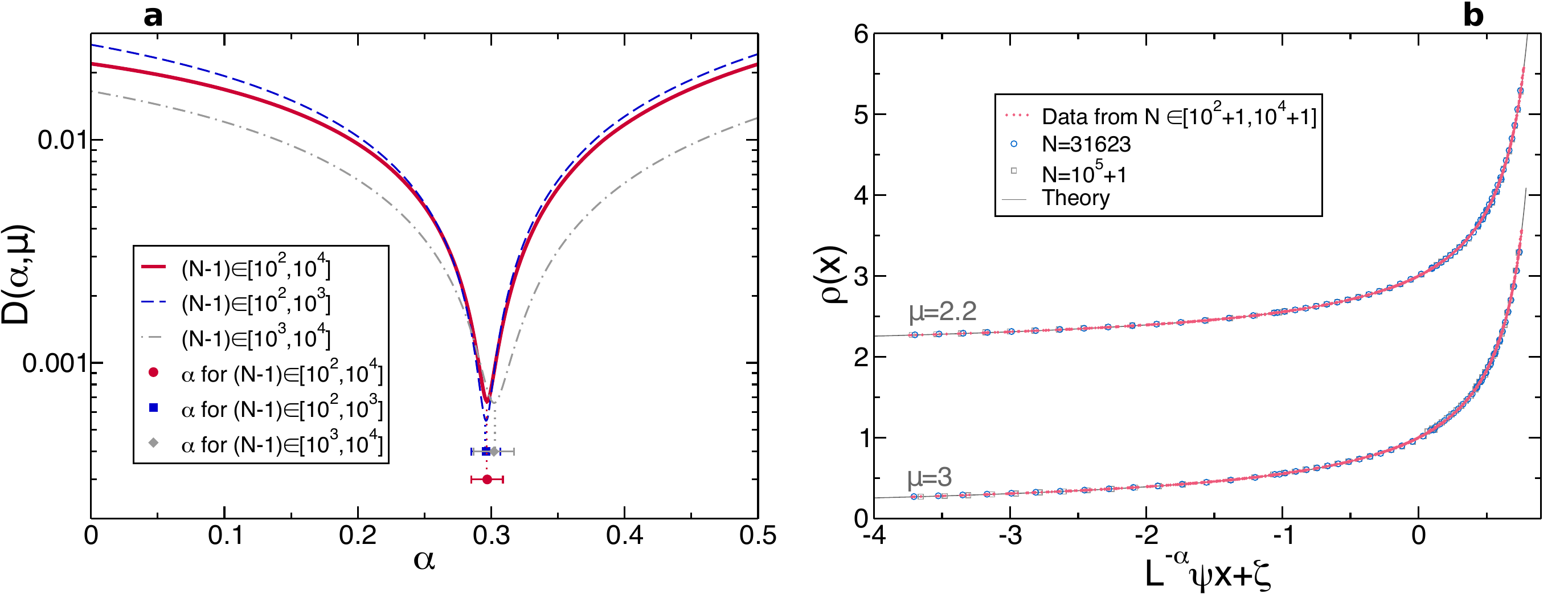}
}
\vspace{-0.3cm}
\caption{\small  {\bf Ruling out finite-size corrections.} (a) Distance metric $D(\alpha,\mu)$ for $\mu=3$ as a function of $\alpha$ when considering all data, $N\in[10^2+1,10^4+1]$, $T_0\in[2,20]$, $\eta\in[0.5,3]$ (full line), and when $N$ is restricted to small $N\in[10^2+1,10^3+1]$ (dashed line) or large $N\in[10^3+1,10^4+1]$ (dot-dashed line). Notice the logarithmic scale in the $y$-axis. The points and the errorbars below represent the estimated value of the anomaly exponent $\alpha$ in each case. Clearly, values of $\alpha$ obtained from the restricted sets in $N$ are fully compatible between them and with the previous result using the combined sets, all points lying well within the errorbars. Note that the distance curve for large $N$ is slightly wider than the small-$N$ curve due to the somewhat larger uncertainties accompanying data for large $N$, a direct result of the strong growth of relaxation and correlation times with $N$, see \rr{Supplementary Fig. S4} and related discussion. 
(b) Collapse of density profiles for $\mu=2.2$ (top) and $\mu=3$ (bottom) obtained by using the measured anomaly exponent $\alpha$ in each case, see \rr{Supplementary Table S1}. Small points correspond to the scaling collapse obtained for $N\in[10^2+1,10^4+1]$, $T_0\in[2,20]$, and $\eta\in[0.5,3]$, while bigger points correspond to additional results obtained from extensive simulations for larger system sizes, namely $N=31623$ ($\bigcirc$) and $N=10^5+1$ ($\Box$), with $T_0=20$ and $\eta\in[0.5,3]$. The line stands for the theoretical prediction, and the master curve for $\mu=2.2$ has been shifted vertically for better comparison.
}
\label{fig5}
\end{figure*}

\rr{
At this point it is worth emphasizing that standard linear response methods to measure the heat conductivity typically yield an \emph{effective} anomaly exponent in $1d$ which changes appreciably with the system size, $\tilde{\gamma}(N)$, slowly converging to the asymptotic value $\gamma$ at very large $N$ \cite{chen14a}, see Section II.C in the Supplementary information. A natural question is hence whether the new anomaly exponent $\alpha(\mu)$ measured with the novel scaling method introduced here exhibits similar finite-size corrections. A first clue that this is not the case is that, for $N\in[10^2+1,10^4+1]$, a slight change in the anomaly exponent measured with our scaling method completely destroys the observed collapse, see Fig. \ref{fig2}.b, while the effective anomaly exponent measured with standard methods varies widely with $N$ in the same $N$-range, e.g. $\tilde{\gamma}(N)\in[0.25,0.5]$, see Fig. 3.b in Ref. \cite{chen14a}. In any case, in order to test quantitatively this idea, we divided our original data into two different subsets, one for small $N\in[10^2+1,10^3+1]$ and another one for large $N\in[10^3+1,10^4+1]$. In this way both data subsets have the same amount of points, thus avoiding possible sampling issues. Next, we perform our scaling analysis on both subsets and obtain the collapse distance metric $D(\alpha,\mu)$ as a function of $\alpha$ in each case. In both cases, small $N$ vs large $N$, this function exhibits a pronounced minimum in $\alpha$ for each $\mu$, and these minima identify the anomaly exponent as measured in each subset. Fig. \ref{fig5}.a shows the results of this analysis for mass ratio $\mu=3$, and the conclusion is clearcut: the anomaly exponents measured from the small-$N$ and large-$N$ subsets are fully compatible between them and with our previous measurement based on all $N\in[10^2+1,10^4+1]$, so no significant, systematic variation of the anomaly exponent with the system size is found beyond the stringent errorbars of our measurements. We found similar results for all other $\mu$'s.
}

\rr{
To further test the robustness of the measured anomaly exponents against order-of-magnitude changes in the system size, we also studied the steady-state heat transport in the diatomic hard-point fluid for $N=31623$ and $N=10^5+1$, i.e. one order of magnitud beyond our previous simulations. The scale of these simulations is so large that we had to restrict the region of parameter space explored. In particular, we perform simulations of the aforementioned values of $N$ for a large temperature gradient given by $T_0=20$, global densities $\eta=0.5, 1, 2, 3$, and two intermediate mass ratios $\mu=3$ and $\mu=2.2$ for which relaxation (and correlation) timescales are somewhat shorter (note that for both small and large mass ratios the fluid's relaxation and correlation times increase drastically \cite{hurtado06c,hurtado06d}). Fig. \ref{fig5}.b shows the collapse of density profiles for $\mu=2.2$ and $\mu=3$ obtained by using the measured anomaly exponent $\alpha(\mu)$ in each case, namely $\alpha(\mu=2.2)=0.308$ and $\alpha(\mu=3)=0.297$, see Supplementary Table S1, once the new data for $N=31623$ and $N=10^5+1$ have been added. In all cases the excellent collapse of all data for $N\in[10^2+1,10^5+1]$, i.e. across three orders of magnitude in the system size, strongly confirms the validity of the measured (non-universal) exponents in the large-$N$ limit. Similar excellent collapses are also obtained for temperature profiles. Moreover, if a different anomaly exponent is used in the previous scaling plots (e.g. $\alpha=1/3$) no good collapse is obtained, as observed in e.g. Fig. \ref{fig2}.b above, even if we restrict the plot to the largest values of $N$. These observations thus discard the possibility of a running anomaly exponent (at least within our stringent precision limits), demonstrating the robustness of the anomaly exponent $\alpha$ against order-of-magnitude changes in the system size and hence strengthening our conclusions. 
}

\begin{figure}
\vspace{-0.3cm}
\centerline{\includegraphics[width=9cm]{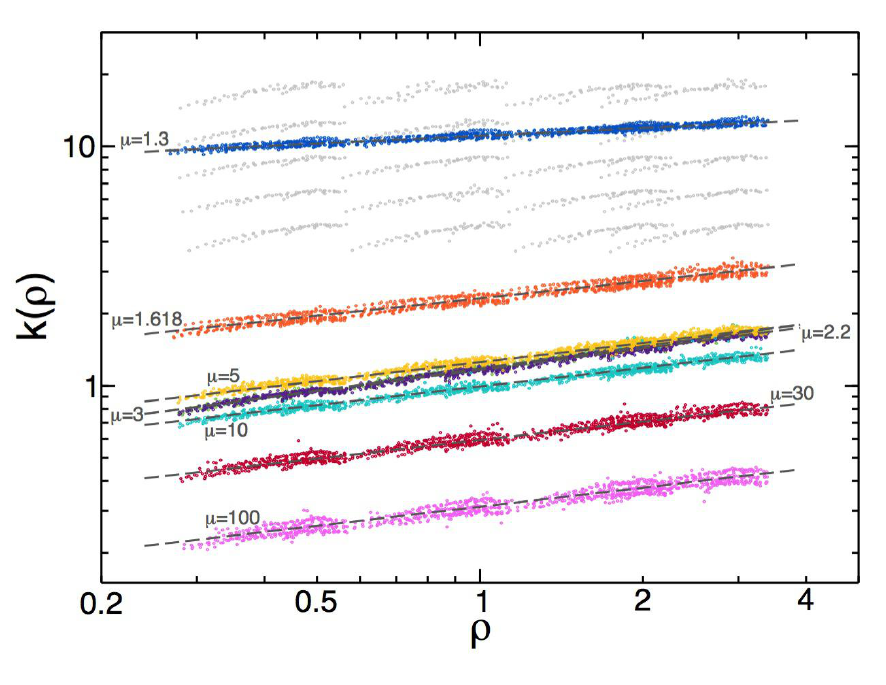}}
\vspace{-0.3cm}
\caption{\small  
{\bf Density dependence of heat conductivity as captured by $k(\rho)$.} Light gray points show the curves obtained for $\mu=3$ before scaling data by $L^{-\alpha}$ along the $y$-axis, while dark color curves show the scaled curves for each $\mu$. A power-law behavior is apparent in all cases. Dashed lines are power-law fits to the data, see main text and Supplementary Table S1. 
}
\label{fig9}
\end{figure}

We next focus on the density dependence of the heat conductivity $\kappa_L(\rho,T)=L^\alpha \sqrt{T/m} \, k(\rho)$. Interestingly, the dynamics of $1d$ hard-point fluids remains invariant under different scalings (of temperature, velocities, space, mass, etc.) \cite{dhar08a}. Using such invariance, it is easy to show rigorously that $\kappa_L(\rho,T)=\sqrt{T/m} f(N,\mu)$, with $f$ some adimensional function of $N$ and $\mu$. This in turn implies, via dimensional analysis, that necessarily $k(\rho)=a\rho^\alpha$, with $a$ some constant. This is fully confirmed in local measurements of the density dependence of the heat conductivity, from which we determine $a=a(\mu)$. \rr{Indeed, one can easily show from Eq. (\ref{conductL}) that $k(\rho)=J \sqrt{m}[L^{\alpha}\sqrt{T(x)}|T'(x)|]^{-1}$, so for each set $(N,T_0,\eta)$ and fixed $\mu$ we performed discrete derivatives of the measured bulk temperature profile to evaluate $T'(x)$ and plotted the previous expression, with $J$ measured in each case, as a function of the associated $\rho(x)$. Fig. \ref{fig9} shows the curves $k(\rho)$ so obtained for different $\mu$, which display the best collapse when the measured exponent $\alpha(\mu)$ is used. Interestingly the resulting scaling functions, though somewhat noisy due to discretization effects, exhibit a clear power-law behavior, $k(\rho)=a\rho^{\beta}$, and the fitted exponent is fully compatible in all cases with the measured anomaly exponent, $\beta= \alpha(\mu)$, see Fig. \ref{fig3}.a above and Supplementary Table S1. These measurements thus prove in an independent way that the density dependence of the heat conductivity of the $1d$ diatomic hard-point gas does reflect the transport anomaly.}

\rr{The above} observation \rr{that $k(\rho)=a\rho^\alpha$} opens the door to a full solution of the macroscopic heat transport problem for this model, see Section I of the Supplementary information. In particular, the universal master curve $F(u)$ of Eq. (\ref{scale}) is 
\be
F(u) =\left(1- \frac{u}{\nu^*}\right)^{\frac{2}{2\alpha-3}}
\label{masterC}
\ee
with $\nu^*\equiv a/(\frac{3}{2}-\alpha)$. This master curve depends on $\mu$ through the mass ratio dependence of $\alpha$ and $a$. Fig. \ref{fig3}.b displays the predicted master curves, with the only input of the measured $\alpha(\mu)$ and $a(\mu)$, and the agreement with collapsed data is stunning \rr{in all cases}. Closed forms for temperature profiles follow as
\be
T(x)=\left[ T_0^{\frac{3}{2}-\alpha} - \frac{J\sqrt{m}}{\nu^*P^\alpha}L^{-\alpha} x  \right]^{\frac{2}{3-2\alpha}} \, , \quad x\in[0,L] \, ,
\label{temperature}
\ee
with density profiles given as $\rho(x)=P/T(x)$, and $P$ and $J$ simply written in terms of external parameters $T_0$, $T_L$, $\eta$, and $L$, \rr{see Supplementary information, Section I}. Note that this novel macroscopic solution is \rr{fully} compatible with the known scaling symmetries of $1d$ hard-point fluids \cite{dhar08a}. Interestingly, the master curve $F(u)$ exhibits a vertical asymptote at $u=\nu^*$, \rr{see Eq. (\ref{masterC})}, implying the existence of a bound on the scaled current in terms of pressure, 
\rr{
\ben
L^{1-\alpha} \psi &\le&  \psi^*\equiv \nu^* \left(\frac{T_0}{P} \right)^{3/2-\alpha}  \Rightarrow \nonumber \\
&\Rightarrow& L^{1-\alpha} J\le \nu^* T_0^{3/2-\alpha} \frac{P^\alpha}{\sqrt{m}} \quad  \forall~ T_0, T_L, \eta, L \, .
\label{Jbound}
\een
}

\begin{figure*}[t]
\vspace{-0.3cm}
\centerline{
\includegraphics[width=18cm]{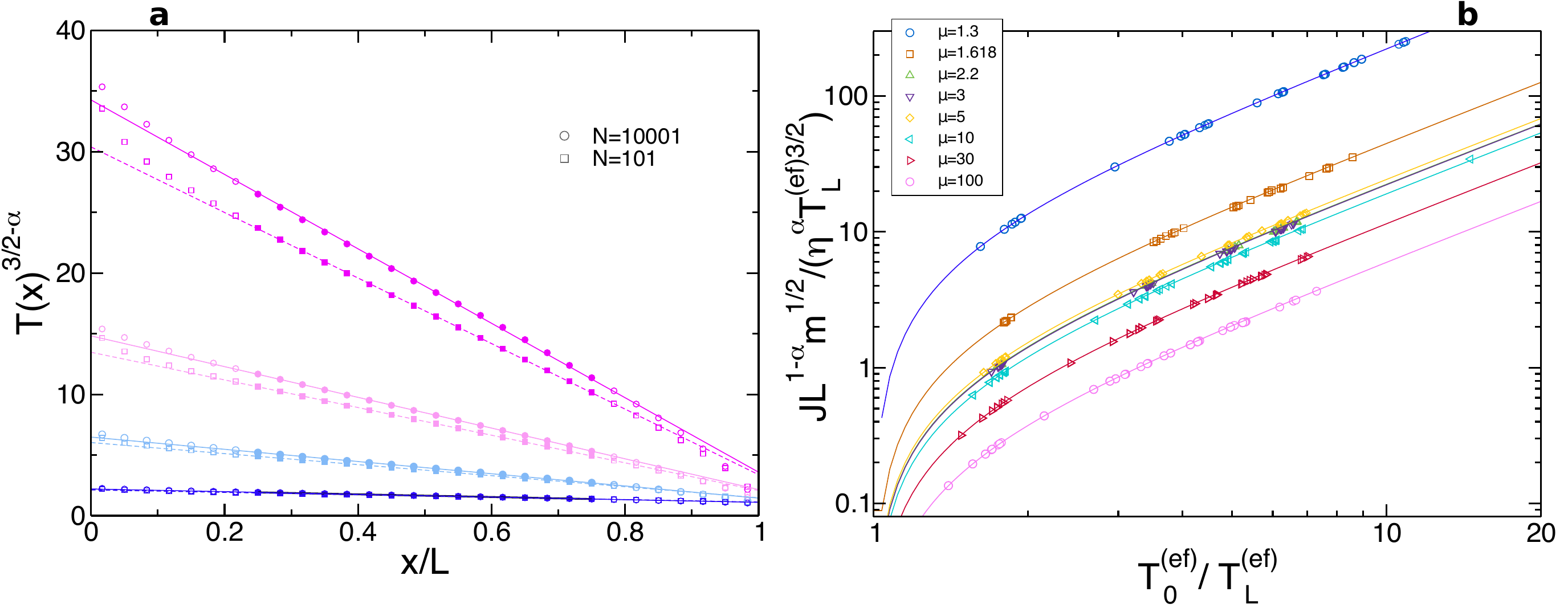}
}
\vspace{-0.3cm}
\caption{\small  {\bf Testing additional predictions.} (a) Measured temperature profiles to the power $(3/2-\alpha)$ vs $x$, for $\mu=3$, $\eta=1$, varying $T_0\in[2,20]$ and two different system sizes, $N=101$ ($\Box$) and $N=10001$ ($\bigcirc$). Filled symbols correspond to the bulk, while open symbols signal the boundary layers. Lines have slope $-(3/2-\alpha)JL^{1-\alpha}\sqrt{m}/(a P^\alpha)$, with $J$ and $P$ the measured current and pressure in each case, and the only fitting parameter corresponds to the $y$-intercept, which yields $T_0^{(\text{ef}) 3/2-\alpha}$ in each case. Note that $T_L^{(\text{ef})}$ follows from $T_0^{(\text{ef})}$ and the (fixed) slope. The agreement between lines and data confirm that \emph{bulk} temperature (and density) profiles for any finite $N$ are in fact those of a \emph{macroscopic} diatomic hard-point gas sustaining a current $J$ and a pressure $P$ and subject to some effective $N$-dependent boundary conditions controlled by the boundary layers. For $\mu=3$, recall that $\alpha=0.297 (6)$ and $a=1.1633 (9)$, see Supplementary Table S1.
(b) Test of the macroscopic theory prediction for the heat current, see Eq. (\ref{JJJ2}). For each mass ratio, $JL^{1-\alpha}\sqrt{m}/(\eta^\alpha T_L^{\text{(ef)}\, 3/2})$ is plotted vs $T_0^{\text{(ef)}}/T_L^{\text{(ef)}}$, with $J$ the measured current, and $T_{0,L}^{\text{(ef)}}$ the effective boundary temperatures for bulk profiles measured in each case. The agreement between data (symbols) and theory (lines) is excellent in all cases.
}
\label{fig6}
\end{figure*}

Eq. (\ref{temperature}) \rr{for temperature profiles} can be readily tested against data. \rr{For that we plot $T(x)^{3/2-\alpha}$ vs $x$, with $T(x)$ the measured temperature profiles for each $\mu$, $N$, $\eta$ and $T_0$. This is predicted to be a straight line with slope $-(3/2-\alpha)JL^{1-\alpha}\sqrt{m}/(a P^\alpha)$, with $J$ and $P$ the measured current and pressure, respectively. Such linear dependence is confirmed for bulk temperature profiles in all cases (similar results hold also for density profiles), with the correct slope but with effective boundary temperatures (obtained from the $y$-intercept of the line) slightly different from the thermal wall temperatures in each case, $T_{0,L}^{\text{(ef)}}(N)\ne T_{0,L}$. Fig. \ref{fig6}.a shows an example of this test for $\mu=3$, $\eta=1$, varying $T_0\in[2,20]$ and two different system sizes, $N=101$ (small) and $N=10001$ (large), with excellent agreement in all cases.} This shows that the measured \emph{bulk} temperature (and density) profiles for any finite $N$ are in fact those of a \emph{macroscopic} diatomic hard-point gas \rr{sustaining a current $J$ and a pressure $P$ and} obeying Eqs. (\ref{fourierL})\rr{-(\ref{conductL})}, but subject to some effective $N$-dependent boundary conditions controlled by the boundary layers. Indeed, the striking collapse of data and the agreement with the macroscopic master curve in \rr{Fig. \ref{fig3}}.b strongly support this conclusion. This is a manifestation of the bulk-boundary decoupling phenomenon already reported in hard disks out of equilibrium \cite{pozo15b}, which enforces the macroscopic laws on the bulk of the finite-sized fluid. 

The effective boundary temperatures converge toward $T_{0,L}$ as $N$ increases, but at an exceedingly slow rate, \rr{$[T_0-T_0^{\text{(ef)}}(N)]/T_0 \sim \Lambda/\sqrt{\log N}$ (see Supplementary Fig. S5), with $\Lambda$ some amplitude,} and this explains the persistent finite-size corrections found in the effective anomaly exponents measured with traditional linear response methods. 
\rr{
Indeed, these methods approximate the heat conductivity as $\tilde{\kappa}\approx J L /\Delta T$, with $\Delta T=T_0-T_L$, and find that the so-defined empirical conductivity diverges as $\tilde{\kappa}\sim N^{\tilde{\gamma}(N)}$ in $1d$, with $\tilde{\gamma}(N)$ an \emph{effective} anomaly exponent which exhibits itself persistent finite-size corrections \cite{lepri03a,dhar08a,lepri16a}. Noting that the real temperature gradient driving the bulk fluid to sustain a current $J$ is $\Delta T^{\text{(ef)}}=T_0^{\text{(ef)}}-T_L^{\text{(ef)}}$ and taking into account the strong finite-size corrections affecting the boundary effective temperatures, it is easy to show (see Section II.C of the Supplementary information) that    
\be
\tilde{\gamma}(N) = \gamma + \frac{ \log\left(1-\frac{\Lambda}{\sqrt{\log N}}\right) }{ \log N} \, , 
\label{alphaef}
\ee
so the effective anomaly exponent $\tilde{\gamma}(N)$ measured from the empirical conductivity $\tilde{\kappa}$ converges at an exceedingly slow rate toward the correct, asymptotic anomaly exponent $\gamma$, in a way that closely resembles actual measurements, see e.g. Ref. \cite{chen14a}. This confirms that the slowly-decaying (and artificial) \emph{boundary} finite-size corrections associated to the boundary layers are responsible of the strong, persistent finite-size deviations affecting the effective anomaly exponent measured with the standard linear response method. Moreover, as our scaling method is independent of the boundary temperatures driving the system out of equilibrium, this explains why our results for the anomaly exponent $\alpha$ (that we conjecture is equal to $\gamma$) are free of these persistent finite-size corrections.
}

Finally, our macroscopic theory also offers a precise prediction for the heat current, see the Supplementary information, Section I. In particular, it predicts that \rr{$JL^{1-\alpha}\sqrt{m}/(\eta^\alpha T_L^{3/2}) = h_\alpha (T_0/T_L)$, with $h_\alpha(z)$ a well-defined function
\be
h_\alpha(z)\equiv  a\frac{(\frac{1}{2}-\alpha)^\alpha}{(\frac{3}{2}-\alpha)^{1+\alpha}} \, \frac{(z^{3/2-\alpha}-1)^{1+\alpha}}{(z^{1/2-\alpha}-1)^{\alpha}} \, . 
\label{JJJ2}
\ee
}This prediction can be tested against data using the effective boundary temperatures $T_{0,L}^{\text{(ef)}}$ measured above, see Fig. \ref{fig6}.b, and the agreement is excellent $\forall N, T_0, \eta$ for each $\mu$.

\section{Discussion}

Some comments are now in order. The excellent collapse of our data confirms that \rr{anomalous} Fourier's law (\ref{fourierL}) holds in this model with a well-defined (albeit size-dependent) conductivity functional $\kappa_L(\rho,T)=a(\rho L)^\alpha\sqrt{T/m}$. This is true even for finite $N$ (as small as ${\cal O}(10^2)$!) and under large temperature gradients, extending the range of validity of anomalous Fourier's law deep into the nonlinear regime and evidencing the absence of higher-order (Burnett-like) corrections in $1d$ \cite{pozo15b}. 

In addition, we provide strong evidences supporting the breakdown of universality in anomalous Fourier's law for $1d$ momentum-conserving systems \cite{xiong12a}. \rr{In particular, we show with high accuracy that the new anomaly exponent $\alpha$ for the heat conductivity of the $1d$ diatomic hard-point fluid depends on the mass ratio $\mu$ between neighboring particles. This clear-cut observation, together with the conjectured equality between the different anomaly exponents, $\alpha=\gamma(=\delta)$, calls into question the universality picture for heat transport based on renormalization-group and mode-coupling calculations \cite{narayan02a,beijeren12a}. Note however that our results do not say anything about or contradict the L\'evy/KPZ universality of the equilibrium time correlators of the \emph{conserved} (hydrodynamic) fields, recently predicted within nonlinear fluctuating hydrodynamics and tested in simulations \cite{mendl13a,spohn14a,das14a,mendl15a}.}

\rr{
Different tests of the universality conjecture for the heat transport anomaly have been performed in the past for the diatomic hard-point fluid using a number of methods, including both nonequilibrium simulations of heat transport in the linear response regime and equilibrium measurements of current time-correlation functions \cite{lepri03a,dhar08a,lepri16a}. All tests confirm the existence of the heat transport anomaly for this model. However, the accuracy of the standard methods to determine the anomaly exponents is severely hampered by the strong finite-size corrections affecting these measurements, making very difficult to discern the breakdown of universality here reported. For instance, determining the heat conductivity via the standard nonequilibrium route leads to a running effective anomaly exponent $\tilde{\gamma}(N)$ which exhibits itself persistent finite-size deviations and poor convergence with $N$ \cite{chen14a}. Our scaling results explain the origin of this extremely slow convergence, which in brief can be traced back to the mixing of the artificial but very strong \emph{boundary} finite-size corrections with the most important \emph{bulk} scaling behavior. Since our collapse procedure is independent of the boundary driving, this explains why our scaling results for the anomaly exponent $\alpha$ are free of these persistent finite-size effects, offering very precise measurements which remain robust across three decades in $N$. On the other hand, the standard equilibrium (Green-Kubo) route to study the anomaly can typically test the \emph{compatibility} of the long-time tail exponent $\delta$ with the universality prediction, but cannot discriminate in most cases the small exponent differences associated to the universality violation here reported. This is particularly relevant for mass ratio $\mu=3$, for which most equilibrium tests have been performed and where our scaling results yield an anomaly exponent close to (but different from) $1/3$, the universality prediction for this model. Therefore it would be desirable to perform standard equilibrium tests also for other mass ratios for which the difference between the universality exponent and the one we measure from scaling are more definite, as e.g. $\mu=10$ for which $\alpha=0.260 (14)$, see Supplementary Table S1. We note however that some recent and very precise simulations of the equilibrium diatomic hard-point fluid for $\mu=3$ and $N=4096$ suggest \cite{mendl15a} an equilibrium anomaly exponent $\delta=0.33 > \alpha(\mu=3)=0.297 (6)$. This apparent discrepancy, which needs further investigation, could mean that the relation between the different anomaly exponents is not as straightforward as conjectured.
}

Which is the origin of the universality breakdown here reported? This violation of universality may hint at the possible existence of hidden \rr{slowly-evolving fields in the diatomic hard-point gas other than the standard (locally-conserved) hydrodynamic ones}. Remarkably, such intriguing behavior has been already reported in the nonequilibrium response of this model to a shock wave excitation \cite{hurtado06a,hurtado05a}, \rr{and suggests that a more convoluted fluctuating hydrodynamics description (including the additional slow fields, as in granular fluids \cite{dufty11a}) may be needed to understand anomalous transport in this model}. Moreover, as recently put forward \cite{popkov15a}, the existence of further \rr{slowly-evolving fields} may give rise to an infinite discrete (Fibonacci) family of anomaly exponents that can coexist in different regions of parameter space for a given model \cite{popkov15a}, \rr{changing from one value to another as a control parameter is varied}, a behavior reminiscent of our results. 

The question remains as to how to reconcile the local nature of Fourier's law with the non-local $L^\alpha$-term in $\kappa_L(\rho,T)$.  Our data suggest that this could be achieved in a nonlinear fluctuating hydrodynamics description of the problem derived via an anomalous, non-diffusive hydrodynamic scaling of microscopic spatiotemporal variables, $x\to x/L^{1-\alpha}$ and $t\to t/L^{2-3\alpha}$. We also mention that recent results suggest yet another mesoscopic description of anomalous transport in $1d$ in terms of fractional diffusion equations and/or heat carriers with L\'evy-walk statistics \cite{cipriani05a,dhar13a,bernardin16a,jara15a}. As far as we know, this description does not seem compatible with the scaling and data collapses observed in this work. Finally, it would be interesting to apply the scaling method here developed to other paradigmatic models of heat transport in low dimensions, as e.g. the Fermi-Pasta-Ulam model of anharmonic oscillators and the hard-square or -shoulder potentials \cite{bonetto00a,lepri03a,dhar08a,lepri16a}, where the reported universality breakdown can be further investigated. The role of conservative noise \cite{bernardin16a,jara15a} as a smoothing mechanism to get rid of non-hydrodynamic, hidden \rr{slow fields} should be also investigated.

\section*{Acknowledgments}
We thank H. van Beijeren, A. Dhar, J.L. Lebowitz, J.J. del Pozo, and H. Spohn for useful discussions. Financial support from Spanish project FIS2013-43201-P (MINECO), University of Granada, Junta de Andaluc\'{\i}a project P09-FQM4682 and GENIL PYR-2014-13 project is acknowledged.

\section*{Author contributions}
P.I.H. conceived the project, performed the numerical simulations and prepared the figures. P.I.H. and P.L.G. carried out the calculations and analyzed the data. P.I.H. wrote the main manuscript text. All authors reviewed the manuscript.

\section*{Additional information}
{\bf Competing financial interests:} The authors declare no competing financial interests.

\section*{Supplementary Information}
\appendix

\section{Scaling in Fourier's law}

In this Section we will show that the stationary density and temperature profiles of the $1d$ diatomic hard-point fluid driven out of equilibrium by an arbitrary temperature gradient follow from an universal master curve, provided that three simple hypotheses hold (see below). It will be then trivial to show that the reverse statement also holds true, i.e. that a nonequilibrium $1d$ fluid whose density and temperature profiles collapse onto an universal master curve is bounded to obey the three mentioned properties. These hypotheses are:
\begin{itemize}

\item[(i)] {\bf Fourier's law:} 
In the steady state, the nonequilibrium fluid sustains a non-vanishing heat current $J$ proportional to the temperature gradient
\be
J=-\kappa_L(\rho,T)\frac{dT(x)}{dx}\, , \quad x\in[0,L] \, ,
\label{fourierLa}
\ee
with $\kappa_L(\rho,T)$ a well-defined local conductivity functional which may depend on $L$. 

\begin{figure*}
\vspace{-0.3cm}
\centerline{\includegraphics[width=15cm]{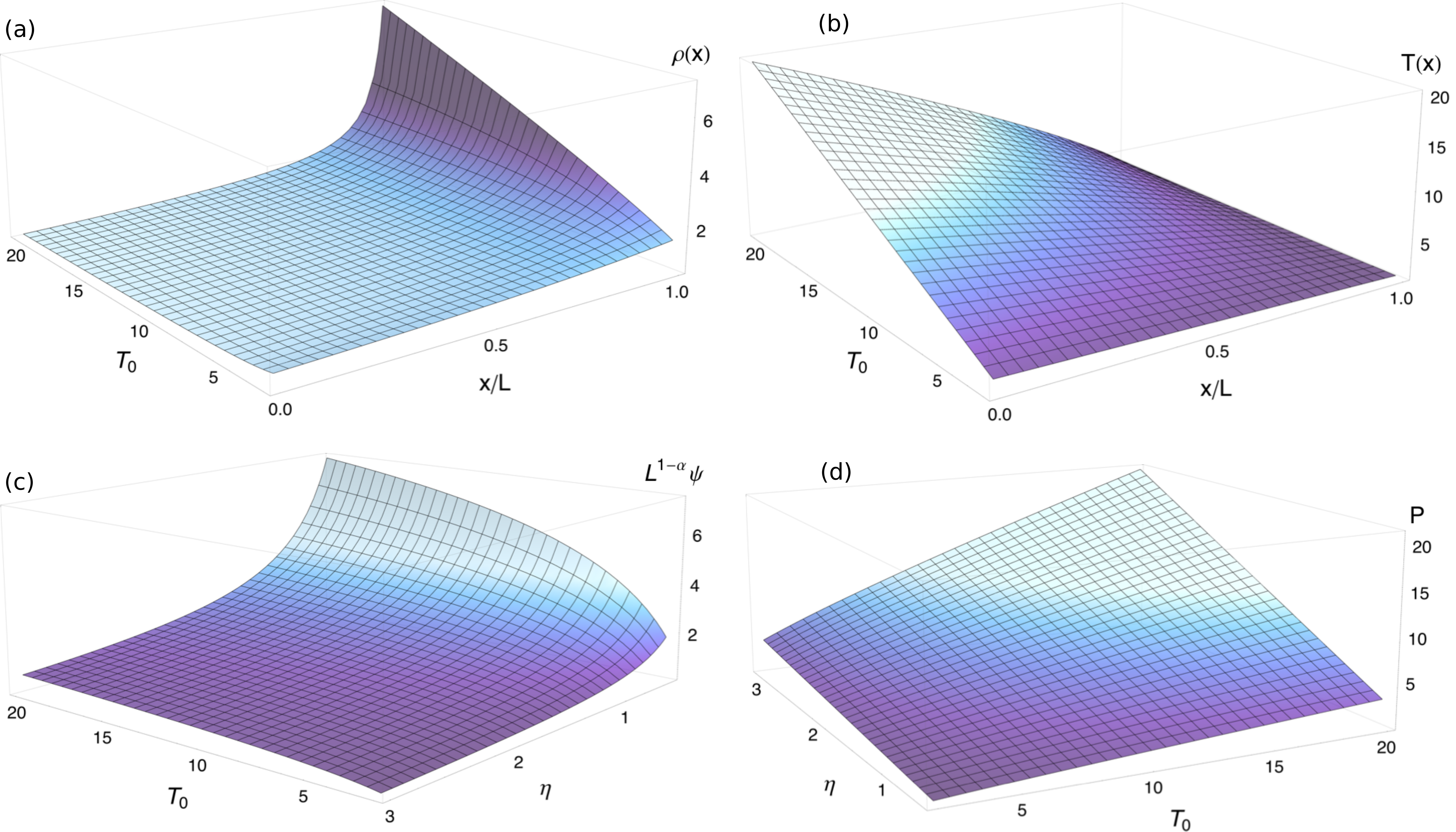}}
\vspace{-0.3cm}
\caption{\small  {\bf Theoretical predictions.} Density (a) and temperature (b) profiles as a function of $T_0$ for $\eta=1$, obtained from the full solution of the macroscopic heat transport problem for the $1d$ diatomic hard-point gas, see Eq. (\ref{densitya2}) and the associated discussion. Also shown are the $\eta$- and $T_0$-dependence of (c) the scaled reduced current, $L^{1-\alpha} \psi$, and (d) the nonequilibrium fluid's pressure $P$, see Eqs. (\ref{PPP})-(\ref{JJJ}). All these curves are for a mass ratio $\mu=3$, for which $\alpha=0.297(6)$ and $a=1.1633(9)$, see Table \ref{tab} below where all measured anomaly exponents for different $\mu$'s can be found.
}
\label{fig6A}
\end{figure*}

\item[(ii)] {\bf Macroscopic local equilibrium (MLE):} This amounts to assume that local thermodynamic equilibrium holds at the macroscopic level, in the sense that the local density and temperature are related by the \emph{equilibrium} equation of state (EoS) \cite{pozo15aA}. For the $1d$ diatomic hard-point gas studied in this paper, it is simply the ideal gas EoS
\be
P=\rho T \, ,
\label{EoSa}
\ee
with $P$ the fluid's pressure.

\item[(iii)] {\bf Heat conductivity scaling:} Due to the homogeneity of the interaction potential, the heat conductivity of the $1d$ diatomic hard-point gas exhibits a well-known density temperature separability \cite{pozo15bA}. Moreover, standard dimensional analysis arguments show that $\kappa\propto \sqrt{T/m}$ \cite{pozo15bA}, and the known dimensional anomaly for transport implies in turn that $\kappa\propto L^\alpha$ at leading order. We now raise these arguments to a formal scaling ansatz 
\be
\kappa_L(\rho,T)=L^\alpha \sqrt{T/m} k(\rho) \, ,
\label{kappa}
\ee
with $k(\rho)$ a function solely dependent on density. Note that this ansatz discards possible subleading corrections in $L$. 
\end{itemize}

We may now use the MLE property (ii) and the conductivity scaling ansatz (iii) to write Fourier's law in terms only of the density field. In particular, using the EoS to write $T(x)=P/\rho(x)$, we obtain
\be
\frac{J\sqrt{m}}{P^{3/2}} L^{-\alpha} = G'(\rho) \frac{d\rho}{dx} = \frac{dG(\rho)}{dx} \, ,
\label{JQcte}
\ee
where $G'(\rho)\equiv k(\rho) \rho^{-5/2}$ and $'$ denotes derivative with respect to the argument. This equation, together with the boundary conditions for the density field, $\rho(x=0,L)=\rho_{0,L}$, which can be inferred from the constraints
\ben
\frac{T_0}{T_L} &=& \frac{\rho_L}{\rho_0}  \, , \label{rho0L1} \\ 
\eta =  \frac{1}{L}\int_0^L \rho(x) dx &=& \frac{\displaystyle \int_{\rho_0}^{\rho_L} \rho \, G'(\rho) d\rho}{\displaystyle G(\rho_L)-G(\rho_0)} \, ,
\label{rho0L}
\een
completely define the macroscopic problem in terms of $\rho(x)$. Note that the externally controlled parameters in the problem are the temperatures of the boundary reservoirs, $T_{0,L}$, and the global number density $\eta$. The pressure and the heat current can be now obtained as $P=T_0 \, \rho_0$ and $J=P^{3/2}[G(\rho_L)-G(\rho_0)]/(L^{1-\alpha}\sqrt{m})$.

A simple yet striking consequence of hypotheses (i)-(iii) can be now directly inferred from Eq. (\ref{JQcte}). In fact, as both $J$ and $P$ are state-dependent constants, this immediately implies that $G[\rho(x)]=\psi L^{-\alpha} x + \zeta$, i.e. $G[\rho(x)]$ is a linear function of position $x\in[0,L]$, with $\psi=J\sqrt{m}/P^{3/2}$ the reduced current and $\zeta=G(\rho_0)$ a constant. Equivalently, 
\be
\rho(x)=F\left(\frac{\psi}{L^\alpha} x + \zeta\right) \, ,
\label{densitya}
\ee
where we have assumed that the function $G(\rho)$ has a well-defined inverse $F(u)\equiv G^{-1}(u)$. This assumption seems reasonable as steady density profiles are typically well behaved and readily measurable in simulations and experiments, see e.g. Fig. 1 in the main text. Therefore, according to Eq. (\ref{densitya}), there exists a single universal master curve $F(u)$ from which any steady state density profile follows after a linear spatial scaling $x=L^\alpha (u-\zeta)/\psi$. This scaling behavior is automatically transferred to temperature profiles via the local EoS $P=\rho(x) T(x)$, so 
\be
\frac{T(x)}{P}=\frac{\displaystyle 1}{\displaystyle F\left(\frac{\psi}{L^\alpha} x + \zeta\right)} \, .
\label{temperaturea}
\ee
These scaling laws are independent of the global density $\eta$ or the nonequilibrium driving defined by the baths temperatures $T_0$ and $T_L$, depending exclusively on the function $k(\rho)$ controlling the fluid's heat conductivity. Alternatively, Eq. (\ref{densitya}) implies that any measured steady density profile can be collapsed onto the universal master curve $F(u)$ by scaling space by the scaled reduced current $L^{-\alpha}J\sqrt{m}/P^{3/2}$ measured in each case and shifting the resulting profile an arbitrary constant $\zeta$ (similarly for temperature profiles). This suggests a simple scaling method to obtain the universal master curves in simulations or experiments, a procedure that we implement in the main text. Note that these results are not limited to the $1d$ diatomic hard-point gas; equivalent results hold for general $d$-dimensional (non-critical) fluids driven arbitrarily far from equilibrium, see Ref. \cite{pozo15bA} for a proof.

Proving the reverse statement, i.e. that a $1d$ fluid obeying Eqs. (\ref{densitya})-(\ref{temperaturea}) does fulfill also properties (i)-(iii) above, is now trivial. In particular, the MLE property (ii) is automatically satisfied by construction. Moreover, inverting the scaling in (\ref{densitya}) to obtain $G[\rho(x)]$ and differentiating this functional with respect to $x$ we arrive at $J=-L^{\alpha} \sqrt{T/m} G'(\rho) \rho^{5/2} T'(x)$, where we used that $T(x)=P/\rho(x)$, see Eqs. (\ref{densitya})-(\ref{temperaturea}). This hence proves that properties (i) and (iii) also hold, with a heat conductivity given by Eq. (\ref{kappa}) with $k(\rho)=G'(\rho) \rho^{5/2}$.

The combination of our scaling ansatz for the heat conductivity and well-known dynamical invariances of $1d$ hard point fluids under scaling of different magnitudes (as e.g. temperature, velocities, mass, space, etc.) results in a well-defined density dependence for the heat conductivity, see main text, namely $k(\rho)=a\rho^\alpha$, with $a$ a constant of ${\cal O}(1)$. Such power-law dependence, which reflects the transport anomaly, is fully confirmed in local measurements of the density dependence of $\kappa_L$, see Fig. 5 in the main text, from which we obtain precise estimates of the amplitude $a(\mu)$, see Table \ref{tab}. Such clear-cut observation, together with the scaling formalism described above, allows now for a complete solution of the \emph{macroscopic} transport problem for this model, written in terms of the external control parameters, namely $T_0, T_L, \eta$ and $L$, together with $\alpha$ and $a$. In fact, recalling that $G'(\rho)= k(\rho) \rho^{-5/2}$ we obtain that $G(\rho)=\nu^*(1-\rho^{\alpha-3/2})$, with $\nu^*\equiv a/(\frac{3}{2}-\alpha)$ and where we have chosen an arbitrary constant such that $F(0)=1=G^{-1}(0)$. The universal master curve hence reads
\be
F(u)=(1-\frac{u}{\nu^*})^{\frac{2}{2\alpha-3}} \, .
\label{masterca}
\ee
This prediction is compared with the measured master curves in Fig. 3 (right panel) of the main text, and the agreement is excellent for all mass ratios $\mu$. Eq. (\ref{masterca}) implies in turn that density profiles can be written as
\be
\rho(x)=\left[ \left(\frac{P}{T_0}\right)^{\alpha-\frac{3}{2}} - \frac{\psi}{\nu^*}L^{-\alpha} x  \right]^{\frac{2}{2\alpha-3}} \, , 
\label{densitya2}
\ee
while temperature profiles simply follow from $T(x)=P/\rho(x)$, namely 
\be
T(x)=\left[ T_0^{\frac{3}{2}-\alpha} - \frac{J\sqrt{m}}{\nu^*P^\alpha}L^{-\alpha} x  \right]^{\frac{2}{3-2\alpha}} \, .
\label{temperaturea2}
\ee
The calculation is completed by expressing the heat current $J$ and the pressure $P$ in terms of the external parameters by using Eqs. (\ref{rho0L1})-(\ref{rho0L}) above. This yields
\ben
P &=&   \eta \left(\frac{\frac{1}{2}-\alpha}{\frac{3}{2}-\alpha}\right) \, \left(\frac{T_0^{3/2-\alpha}-T_L^{3/2-\alpha}}{T_0^{1/2-\alpha}-T_L^{1/2-\alpha}}\right)        \, , \label{PPP} \\
J &=&   \frac{a \eta^\alpha (\frac{1}{2}-\alpha)^\alpha}{L^{1-\alpha} \sqrt{m}(\frac{3}{2}-\alpha)^{1+\alpha}} \, \frac{(T_0^{3/2-\alpha}-T_L^{3/2-\alpha})^{1+\alpha}}{(T_0^{1/2-\alpha}-T_L^{1/2-\alpha})^{\alpha}}       \, . \label{JJJ}  
\een
The last equation for the current can be rewritten as $J=\eta^\alpha L^{\alpha-1} m^{-1/2} T_L^{3/2} h_\alpha(T_0/T_L)$, with
\be
h_\alpha(z)\equiv  a\frac{(\frac{1}{2}-\alpha)^\alpha}{(\frac{3}{2}-\alpha)^{1+\alpha}} \, \frac{(z^{3/2-\alpha}-1)^{1+\alpha}}{(z^{1/2-\alpha}-1)^{\alpha}} \, .
\label{JJJ2}
\ee
These predictions are fully confirmed by simulations data, see Fig. 6 in main text and Section II below. As a self-consistent check, note that in the equilibrium limit $T_0\to T_L$ both the pressure and the heat current converge to their expected value, namely $P=\eta T_L$ and $J=0$. Fig. \ref{fig6A} shows the density and temperature profiles predicted for a \emph{macroscopic} diatomic hard-point fluid as a function of $T_0$ for $\eta=1$, as well as the pressure and the scaled reduced current $L^{-\alpha}\psi$ as a function of $T_0$ and $\eta$. These plots are obtained for a particular mass ratio $\mu=3$, for which $\alpha=0.297(6)$ and $a=1.1633(9)$, see Table \ref{tab}, and yield an excellent comparison with simulation data, see Fig. 1 in the main text and Fig. \ref{fig8} in Section II. 

\rowcolors{2}{gray!15}{white}
\begin{table}
\begin{tabular}{|l|l|l|l|}
\hline
\multicolumn{1}{|c|}{$\mu$} & \multicolumn{1}{|c|}{$\alpha$} & \multicolumn{1}{|c|}{$\beta$} & \multicolumn{1}{|c|}{$a$} \\ \hline \hline
1.3   & 0.108 (9)  & 0.109 (1)   & 11.105 (8)   \\
1.618 & 0.242 (23) & 0.2408 (18) & 2.307 (3)   \\
2.2   & 0.308 (5)  & 0.3068 (11) & 1.1765 (9)  \\
3     & 0.297 (6)  & 0.2964 (11) & 1.1633 (9)  \\
5     & 0.266 (11) & 0.2641 (12) & 1.2622 (12) \\
10    & 0.260 (14) & 0.2632 (19) & 0.9874 (14) \\
30    & 0.258 (18) & 0.257 (1)   & 0.5942 (12)   \\
100   & 0.265 (22) & 0.2648 (23) &  0.3095 (5)          \\ \hline  
\end{tabular}
\caption{{\bf Anomaly exponents.} Measured anomaly exponents $\alpha$ and their error for different mass ratios $\mu$, see Fig. 3 in main paper. Also shown are the fitted exponent and amplitude of the power-law density dependence of the conductivity, $k(\rho)=a\rho^{\beta}$, see Fig. 5 in the main text. Notice that in all cases $\beta=\alpha$ within error bars, as predicted.}
\label{tab}
\end{table}

Interestingly, the master curve $F(u)$ obtained above exhibits a vertical asymptote at $u=\nu^*$, see Eq. (\ref{masterca}), and this implies in turn the existence of a maximal scaled reduced current $\psi^*$. Indeed, for the associated density profile to exist in its whole domain $x\in[0,L]$, see Eq. (\ref{densitya2}), the following condition must hold
\be
\psi \le  \frac{\nu^*}{L^{1-\alpha}} \left(\frac{T_0}{P} \right)^{3/2-\alpha} \equiv \frac{\psi^*}{L^{1-\alpha}} \, ,
\label{maximalRC1}
\ee
with $P$ expressed as in Eq. (\ref{PPP}). This defines a maximal scaled reduced current $\psi^*$, such that the scaled current $L^{1-\alpha} J\le \psi^* P^{3/2}/\sqrt{m}=\nu^* T_0^{3/2-\alpha} P^\alpha/\sqrt{m}$ $\forall~ T_0, T_L, \eta, L$, defining an upper bound on the heat current in terms of the nonequilibrium pressure. The maximal scaled reduced current increases monotonously with $T_0$, saturating to an asymptotic value in the $T_0\to \infty$ limit, namely
\be
\psi^* \xrightarrow[T_0\to\infty]{}  \frac{a\, (\frac{3}{2}-\alpha)^{1/2-\alpha}}{[\eta\, (\frac{1}{2}-\alpha)]^{3/2-\alpha}} \, . 
\label{maximalRC2}
\ee
Note however that both $L^{1-\alpha} J$ and $P$ diverge as $T_0\to \infty$, though $\psi^*$ remains finite. 

To end this section, we remark that Eqs. (\ref{masterca})-(\ref{JJJ}) constitute the solution of the \emph{macroscopic} transport problem for this model. A comparison of the predicted density and temperature profiles, see Eqs. (\ref{densitya2})-(\ref{temperaturea2}), with the finite-size data of Fig. 1 in the main text allow us to investigate in the main text the bulk-boundary decoupling phenomenon in detail by quantifying the jump between the effective boundary conditions imposed by the boundary layers on the bulk fluid and the empirical bath temperatures.

\section{Additional results}

In this Section we provide additional data, obtained from our extensive simulations of the $1d$ diatomic hard-point fluid model, which support our conclusions in the main text.

\subsection{Macroscopic local equilibrium, pressure and reduced current}

\begin{figure}
\vspace{-0.3cm}
\centerline{\includegraphics[width=8.5cm]{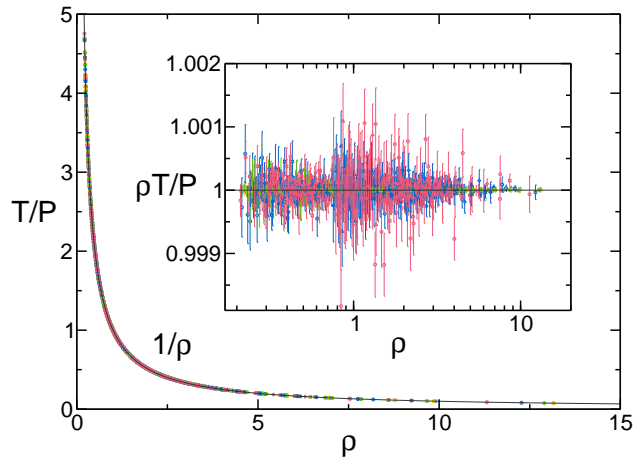}}
\vspace{-0.3cm}
\caption{\small  {\bf Macroscopic local equilibrium.} Measured local reduced temperature, $T(x)/P$, plotted as a function of the associated local density $\rho(x)$ for $\mu=3$ and $\forall \, T_0,\eta, N$, corresponding to all profiles displayed in Fig. 1 of the paper and summing up to 2400 data points from 80 different simulations. An excellent data collapse is obtained which follows with high precision the expected ideal-gas behavior $1/\rho$, plotted as a thin line. Inset: Scaling plot of $\rho(x) T(x)/P$ vs $\rho(x)$ for the same conditions. These data show that macroscopic local equilibrium is a very robust property, even in the presence of strong finite-size corrections on the hydrodynamic profiles.  
}
\label{fig7}
\end{figure}

Our first goal is to test the macroscopic local equilibrium (MLE) property directly from our data. As described above, MLE conjectures that local thermodynamic equilibrium holds at the macroscopic level, in the sense that the stationary density and temperature fields are locally related by the \emph{equilibrium} equation of state (EoS) \cite{pozo15aA}, which for this model is simply the ideal gas EoS,
\be
P=\rho(x) T(x) \, . \nonumber
\label{EoSa2}
\ee
In order to test MLE, we hence take the density and temperature profiles of Fig. 1 of main text measured for $\mu=3$ and different $T_0, N, \eta$, and plot in Fig. \ref{fig7} the local reduced temperature, $T(x)/P$, with $P$ the finite-size pressure measured in each simulation (see below), as a function of the associated local density $\rho(x)$. All data, comprising 2400 points from 80 different simulations for widely different systems sizes, temperature gradients and global densities, collapse onto a single curve which follows with high precision the expected ideal-gas behavior $1/\rho$, see line in Fig. \ref{fig7} and inset therein. Note that, interestingly, the excellent data collapse is maintained also for points within the boundary layers near the thermal walls. Moreover, similar results hold for all mass ratios $\mu$ studied in this paper. In this way, the observed high-precision data collapses confirm the robustness of the MLE property far from equilibrium \cite{pozo15aA}, even in the presence of important finite size effects, validating in an independent manner one of the hypotheses underlying the scaling picture of Section I.

\begin{figure}
\vspace{-0.3cm}
\centerline{\includegraphics[width=8.5cm]{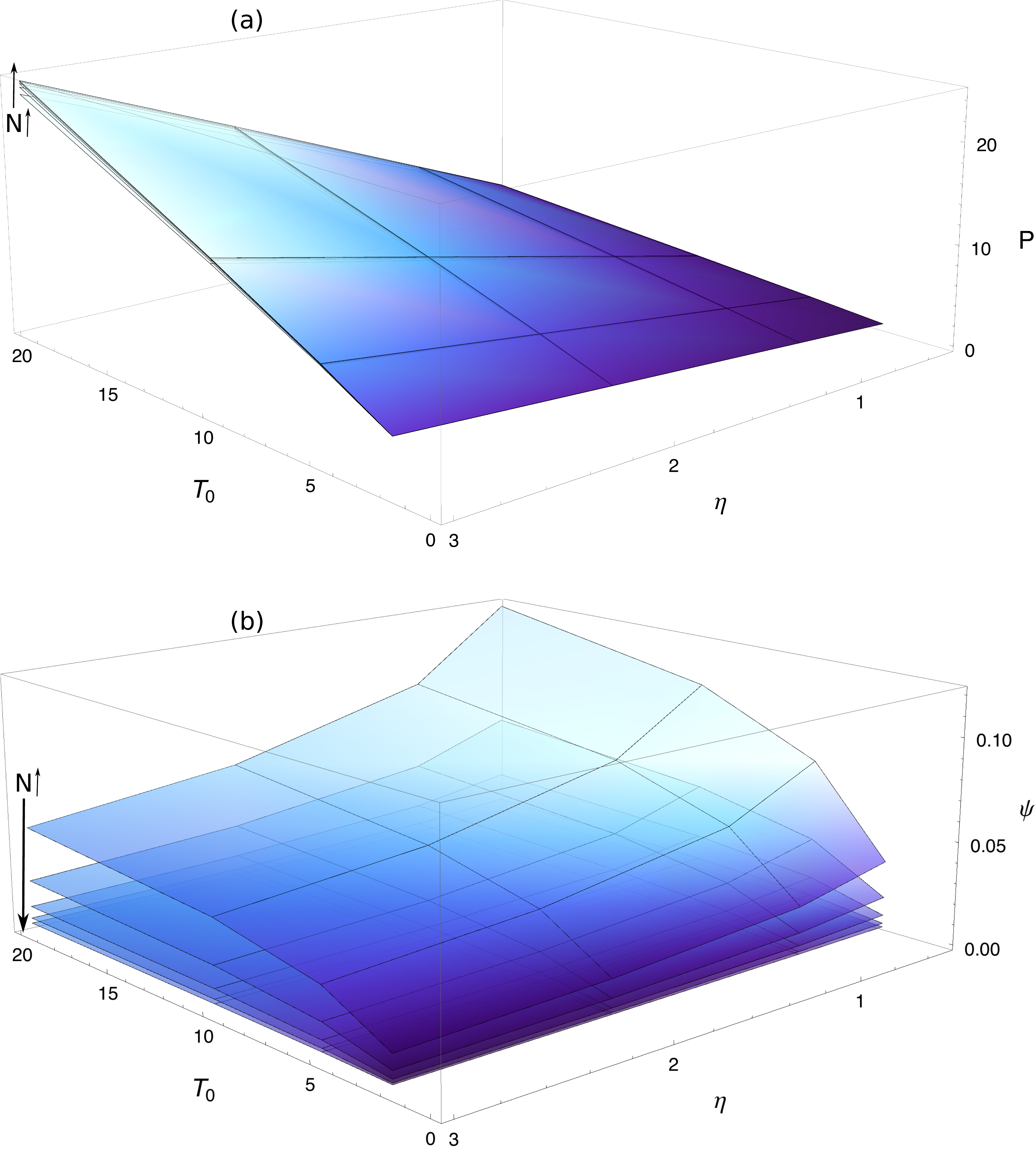}}
\caption{\small  {\bf Pressure and reduced current.} Measured pressure $P$ (a) and reduced current $\psi=J\sqrt{m}/P^{3/2}$ (b) as a function of $T_0$ and $\eta$ for $\mu=3$ and different system sizes $N$. Data here refer to wall observables, though the associated bulk observables yield completely equivalent results.
}
\label{fig8}
\end{figure}

We next focus on the nonequilibrium fluid's pressure $P$ and the heat current $J$ flowing through the system, that we measure both in the bulk and at the thermal walls. These observables are necessary in order to scale the spatial coordinate of the hydrodynamic profiles using the measured reduced current $\psi=J\sqrt{m}/P^{3/2}$ in each case. Fig. \ref{fig8} shows the measured $P$ (a) and $\psi$ (b) as a function of $T_0$ and $\eta$ for $\mu=3$ and different system sizes $N$. These data refer to wall observables, though the associated bulk observables yield completely equivalent results (as otherwise expected). The comparison of these data with our predictions in Section I is excellent, see Fig. \ref{fig6A} above.

\subsection{Slow relaxation in $1d$ transport}

In order to test the robustness of the measured anomaly exponents against order-of-magnitude changes in the system size, we have made a considerable computational effort to characterize the steady-state heat transport in the diatomic hard-point fluid for very large $N$'s, namely $N=31623$ and $N=10^5+1$ (see main text). Of course it would be desirable to go even beyond $N=10^5+1$. However, it is important to note that for such very large values of $N$ obtaining \emph{reliable} results from simulations of $1d$ heat transport is exceedingly difficult. The underlying reason is not only that more particles need more computer time to simulate, but also that relaxation (and correlation) times increase greatly with $N$. This is due to the appearance of current (and momentum) waves in $1d$ which bounce back and forth between the thermal walls at a constant and well-defined speed while they are slowly damped away, a result of the dimensional constraint which strongly suppress local fluctuations. As far as we know, this remarkably slow relaxation mechanism has not been described yet in the literature on $1d$ transport, so we add details about the relaxation process and timescales next.

\begin{figure}[t]
\includegraphics[width=8.5cm,clip]{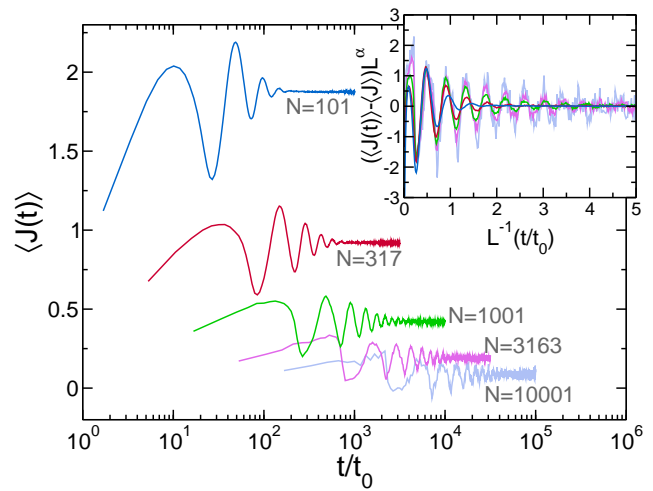}
\vspace{-0.3cm}
\caption{\small  {\bf Slow relaxation in $1d$ transport.} Relaxation of the average instantaneous current as a function of time (measured in units of the mean free time of a heavy particle in a cold environment) for $\mu=3$, $T_0=20$, $\eta=1$, and different values of $N\in[10^2+1,10^4+1]$. Notice the logarithmic scale in the time axis. Relaxation to the nonequilibrium steady state proceeds via a damped current wave with a period which increases linearly with the system length $L$. Inset: Scaling plot of the damped current wave. 
}
\label{fig13}
\end{figure}

In particular, we have performed a large number of relaxation experiments for different values of $N\in[10^2+1,10^4+1]$, for $\mu=3$, $T_0=20$ and $\eta=1$. Initial states for these experiments are randomly drawn from a local equilibrium measure corresponding to the macroscopic density and temperature profiles (obtained from extensive simulations for intermediate system sizes after a long empirical relaxation time). These initial states hence display on average the stationary density and temperature stationary profiles, but lack the weak but long-range correlations which characterize any nonequilibrium steady state (and which are in fact responsible for heat transport). To characterize the relaxation process to the true nonequilibrium steady state, we measured the average instantaneous energy current as a function of time, 
\be
\la J(t)\ra\equiv \frac{1}{2L} \la \sum_{i=1}^N m_i v_i(t)^3 \ra \, , \nonumber
\ee
with the average taken over many different realizations of the relaxation process starting from the random initial states defined previously. Fig. \ref{fig13} shows the relaxation of $\la J(t)\ra$ for different $N$ as a function of time, measured in units of $t_0=\sqrt{M/(2T_L\eta^2)}$, the mean free time of a heavy particle in a cold environment. Note the logarithmic scale in the time axis. From this figure it is clear that the building of the faint but long-range correlations associated to the nonequilibrium steady state proceeds via the formation of a current wave which bounces back and forth between the thermal walls at a constant velocity, while being slowly damped in the nonequilibrium fluid. This damped current wave is accompanied by a similar momentum wave. The period of these waves scales linearly with the system size, and hence the relaxation time to reach the steady state also grows linearly with $L$. In fact, by scaling time by $L^{-1}$ and the excess current by $L^{\alpha}$, with $\alpha(\mu=3)=0.297$, a good collapse is obtained (at least for large $N$), see inset to Fig. \ref{fig13}, which suggest that
\be
\la J(t)\ra = \la J\ra + L^{-\alpha} {\cal G}\left(\frac{t}{Lt_0} \right) \, , \nonumber
\ee
with ${\cal G}(\tau)$ some scaling function. 

The most relevant point here is that relaxation (and correlation) times grow linearly with the system size, making very difficult to obtain good statistics of transport for large enough $N$. In fact, the computer time needed to perform one collision per particle on average in an optimized event-driven molecular dynamics simulation of $N$ particles scales as $(N\log N)\times \tau$ (due to the cost of re-ordering the collision time queue in a heap data structure), with $\tau\sim 10^{-5}\, s$ the typical timescale of an elementary event. As the fluid relaxation and correlation times scale linearly with $N$, the computer time needed to obtain reliable data averages hence scales as $t_\text{sim} \sim n_{\text{exp}}\times (N^2\log N)\times\tau$, with $n_\text{exp}$ the number of measurements to obtain good statistics. For $N=10^5$ and $n_\text{exp}$ in the few hundreds (at least),  we thus have the mind-blowing timescale $t_\text{sim}\sim 10^{8}\, s$, right at the edge of modern-day multiprocessor computer power. For these reasons going beyond $N\sim 10^5$ does not seem feasible at this moment.

\subsection{Finite-size corrections for the effective anomaly exponent measured with standard methods}

One of the most popular methods to measure the heat conductivity of a $1d$ fluid and characterize along the way the associated anomaly exponent consists in setting the model fluid with $N$ particles and density $\eta$ under a small temperature gradient, with fixed wall temperatures $T_{0,L}$, and then increasing $N$ at constant density. For large enough $N$ the overall temperature gradient is small enough so one can \emph{approximate} Fourier's law as
\be
J=-\kappa_L(\rho,T) \frac{dT}{dx}\approx + \tilde{\kappa} \frac{\Delta T}{L} \, , \label{FL}
\ee
with $\Delta T=T_0-T_L$, $J$ the measured current and $L=N/\eta$. In this way, the estimated heat conductivity follows as $\tilde{\kappa}\approx JL/\Delta T$, which is expected to diverge as $N^\gamma$ for \emph{large enough} values of $N$ (though there is no way of knowing a priori which value of $N$ is \emph{large enough}). What it's typically found however in actual, cutting-edge simulations is an effective heat conductivity diverging as $\tilde{\kappa}\sim N^{\tilde{\gamma}(N)}$, with an \emph{effective} anomaly exponent which exhibits itself persistent finite-size corrections \cite{chen14aA}.

Indeed, this approximate method completely neglects the nonlinear density and temperature dependence of the heat conductivity, and by construction it can only yield meaningful results in the limit $N\to\infty$. It is therefore no surprise that the effective anomaly exponent derived within this approach for finite $N$ varies slowly with the system size, as in fact the very definition of $\tilde{\kappa}$ above (and hence $\gamma$) is correct only asymptotically. This weakness in the above definition is reinforced by the fact that, for a given $N$, the heat conductivity measured in this way differs from estimations of $\kappa$ using alternative approaches, as e.g. the also popular Green-Kubo equilibrium method  \cite{chen14aA} (which, by the way, is again exact only in the $N\to\infty$ limit). Moreover, not only the estimated value of $\kappa$ for a given $N$ differs among different approaches, but also its scaling with $N$, and hence the estimation of the anomaly exponent. 

The overall situation is therefore rather unsatisfactory, making extremely difficult to characterize reliably and accurately anomalous Fourier's law in $1d$ with these standard linear response methods. In fact, when measuring $\gamma$ with standard methods one needs to reach huge system sizes, as large as $N\sim 10^5$, to appreciate certain convergence, and even in this case the asymptotic behavior is not yet clearly defined, see e.g. Ref. \cite{chen14aA}.

In contrast with the standard methods described above, our scaling approach takes full advantage of the nonlinear character of the heat conduction problem and provides a fully consistent and highly accurate description of all measured data in a broad range of parameters, including three orders of magnitude in $N\in[10^2+1,10^5+1]$, but also a wide range of temperature gradients (from the linear response regime to the fully nonlinear domain) and densities. The new anomaly exponent $\alpha(\mu)$ that we conjecture equal to $\gamma$ and is measured from the striking collapse of large amounts of data is well-defined, exceptionally robust and does not change with the system size in the broad range explored (see analysis in the main text). This contrasts with the running, effective exponent obtained from the standard linear response methods described above, which varies widely (and far beyond our precision limits) in the same $N$-range, i.e. $\tilde{\gamma}(N)\in[0.25,0.5]$, see e.g. Fig. 3.b in Ref. \cite{chen14aA}. 

\begin{figure}
\vspace{-0.3cm}
\includegraphics[width=8.5cm,clip]{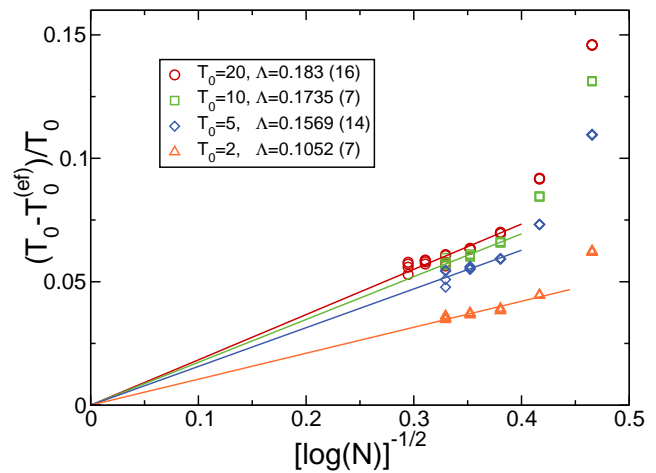}
\vspace{-0.25cm}
\caption{\small {\bf Decay of finite-size corrections for the boundary effective temperature.} The relative temperature gap at the hot thermal wall, defined as $(T_0-T_0^{\text{(ef)}})/T_0$, as a function of $1/\sqrt{\log N}$ for $\mu=3$ and different values of $T_0\in[2,20]$ and $\eta\in[0.5,3]$. Our data are compatible with a linear decay for large enough $N$, $(T_0-T_0^{\text{(ef)}})/T_0 \sim \Lambda/\sqrt{\log N}$, see lines, with $\Lambda$ a small amplitude. This extremely slow, $1/\sqrt{\log N}$ decay of boundary finite-size corrections explains the running effective anomaly exponents previously reported in literature.
}
\label{fig14}
\end{figure}

A natural question now is: why do standard methods measure an effective anomaly exponent which converges so slowly with $N$? The answer to this question lies at the bulk-boundary decoupling phenomenon reported in this work (see main text), i.e. the fact that the bulk of the finite-size nonequilibrium fluid behaves as a macroscopic, infinite system subject to some \emph{effective} boundary temperatures, $T_{0,L}^\text{(ef)}(N)$. Indeed, we measured and characterized the effective $T_{0,L}^{\text{(ef)}}$ $\forall \mu, N, \eta$ and $T_0$ by comparing the measured temperature profiles with the theoretical prediction based on our scaling theory, see left panel of Fig. 6 in the main text and the associated discussion. The effective boundary temperatures so obtained turn out to be $N$-dependent, slightly differing from the wall temperatures $T_{0,L}$ but converging to these values as $N$ increases. Most surprisingly, however, this convergence is exceedingly slow. In fact, Fig. \ref{fig14} shows the measured relative temperature gap at the hot thermal wall, defined as $(T_0-T_0^{\text{(ef)}})/T_0$, as a function of $1/\sqrt{\log N}$ for $\mu=3$ and different values of $T_0\in[2,20]$ and $\eta\in[0.5,3]$, including data for $N=31623$ and $N=10^5+1$ obtained for $T_0=20$. For large enough $N$, namely $N\ge10^3$ (or even $N\ge 317$ for small $T_0$), our data in Fig. \ref{fig14} are compatible with a linear law, hence implying a decay of the form
\be
\frac{T_0-T_0^{\text{(ef)}}(N)}{T_0} \sim \frac{\Lambda}{\sqrt{\log N}} \, , 
\label{Teff}
\ee
with $\Lambda$ a small amplitude. Similar results were obtained for other mass ratios $\mu$ and at the cold boundary. This demonstrates that the effective boundary temperatures the bulk fluid feels (induced by the boundary layers) approach the wall temperatures as $N$ increases at a extremely slow, $\sim 1/\sqrt{\log N}$ rate, and this explains the persistent deviations found in the effective anomaly exponent in literature. More in detail, as explained above standard methods lead to an effective heat conductivity $\tilde{\kappa}\approx JL/\Delta T\sim N^{\tilde{\gamma}}$ for large enough $N$. Using now Eq. (\ref{Teff}) describing the slow decay of boundary finite-size corrections, we have that $T_{0,L}^\text{(ef)}(N) \sim T_{0,L}(1-\Lambda/\sqrt{\log N})$, so that
\be
\Delta T \sim \frac{\displaystyle \Delta T^\text{(ef)}}{\displaystyle 1-\frac{\Lambda}{\sqrt{\log N}}} \, , \nonumber
\ee
with $\Delta T^\text{(ef)}\equiv T_0^\text{(ef)}-T_L^\text{(ef)}$. In this way, for the effective heat conductivity
\be
\tilde{\kappa} \approx \frac{JL}{\Delta T} \sim N^{\tilde{\gamma}} \sim \frac{JL}{\Delta T^\text{(ef)}}\left( 1-\frac{\Lambda}{\sqrt{\log N}} \right) \, . \nonumber
\ee
Noting now that $\Delta T^\text{(ef)}$ is the real temperature gradient felt by the bulk fluid, we thus expect $JL/\Delta T^\text{(ef)}\sim N^\gamma$, so inserting this in the previous equation and taking logarithms we arrive for large $N$ at 
\be
\tilde{\gamma}(N) = \gamma + \frac{ \log\left(1-\frac{\Lambda}{\sqrt{\log N}}\right) }{ \log N} \, , \nonumber
\ee
i.e. an effective anomaly exponent $\tilde{\gamma}(N)$ which converges at a exceedingly slow rate toward the correct, asymptotic anomaly exponent $\gamma$, in a way that closely resembles actual measurements, see e.g. Ref. \cite{chen14aA}. This confirms that the slowly-decaying boundary finite-size corrections associated to the boundary layers are responsible of the strong, persistent finite-size effects affecting the effective anomaly exponent measured with the standard linear response method. Moreover, as our scaling method is independent of the boundary temperatures driving the system out of equilibrium, this explains why our results for the new anomaly exponent $\alpha$, that we conjecture is equal to the true asymptotic exponent $\gamma$, are free of these persistent finite-size corrections.

\section{A metric to quantify data collapse}

In this section we briefly explain the standard metric used in this work to quantify data collapse. This metric is based on the collapse distance first proposed in Ref. \cite{bhattacharjee01aA} and widely used in physics literature, in particular in order to obtain scaling exponents via a distance minimization procedure.  

\begin{figure}
\vspace{-0.3cm}
\centerline{\includegraphics[width=8.5cm]{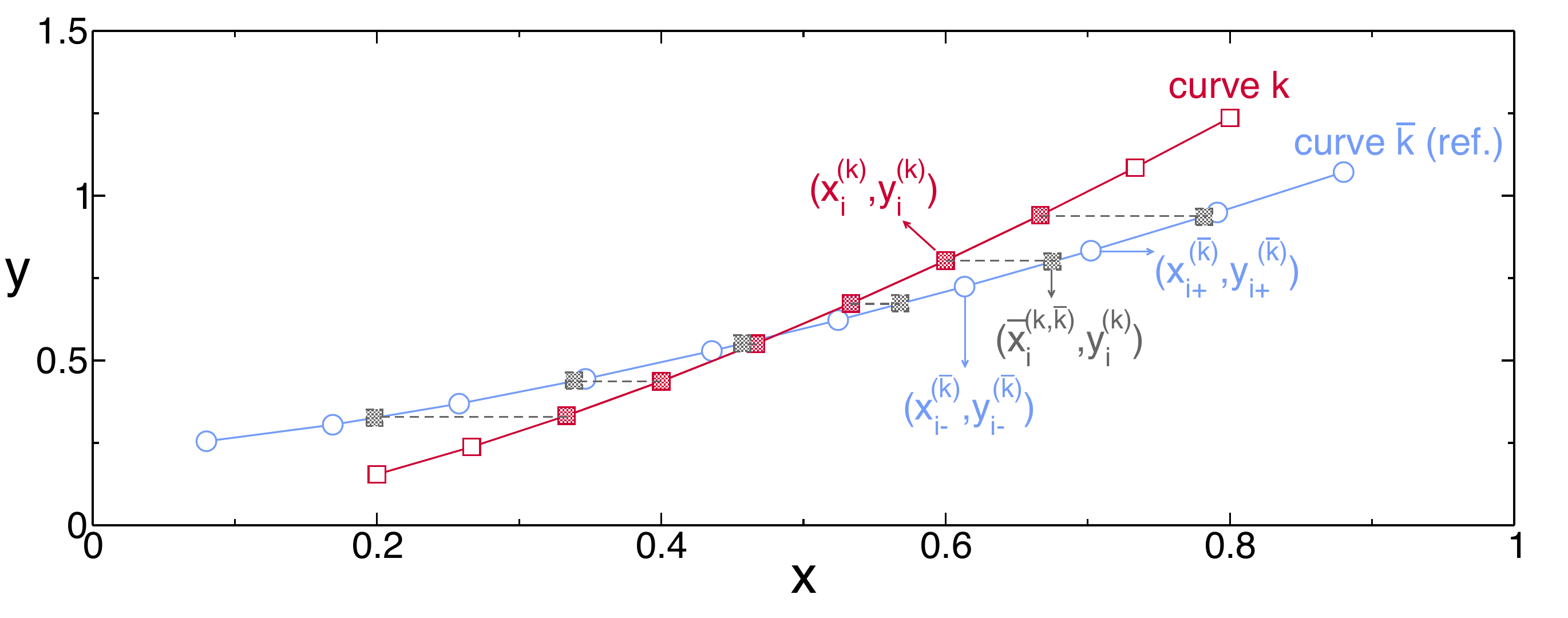}}
\vspace{-0.3cm}
\caption{\small  {\bf A metric to quantify data collapse.} Sketch explaining the metric used to quantify data collapse, see Eq. (\ref{distance}). This metric estimates the distance between a curve $k$ ($\Box$) and the reference curve $\bar{k}$ ($\bigcirc$) by measuring the average distance between each point in $k$ and the interpolated point in $\bar{k}$ with the same $y$-coordinate (gray, shaded squares). Note that we restrict to points in $k$ overlapping with the reference curve $\bar{k}$, see filled squares. The distance corresponds in this example to the average length of the dashed segments.
}
\label{fig11}
\end{figure}

We hence consider a set of $K$ curves, each one containing $M$ points, and we denote this set as $\{ \{(x_i^{(k)},y_i^{(k)}),i\in[1,M] \},k\in[1,K] \}$. The idea is now to choose an arbitrary curve $\bar{k}\in[1,K]$ as reference curve, and proceed to measure the distance of all other curves $k\ne\bar{k}$ to this reference curve along the $x$-direction. For that we measure the distance between each point in $k$ and the interpolated point in $\bar{k}$ with the same $y$-coordinate. In order to do so, we have to restrict to points in $k$ overlapping with the reference curve $\bar{k}$. Note also that we choose to measure distances only along the $x$-direction because the scaling approach developed in this paper only affects the $x$-coordinates of the measured curves, see Section I and Figs. 2 and 3 (right panel) in the main text. Moreover, since the chosen reference curve $\bar{k}$ is completely arbitrary, we repeat this procedure for all curves as reference curve, and average the resulting distances. In this way, our collapse metric is defined as \cite{bhattacharjee01aA}
\be
D\equiv \frac{1}{\ell_{\text{max}} {\cal N}_{\text{overl}}} \sum_{\bar{k}=1}^K \, \sum_{\substack{k=1 \\ k\ne\bar{k}}}^K  \, \sum_{\substack{i=1 \\ i\, \text{overlap} \, \bar{k}}}^M \left\vert x_i^{(k)} - \bar{x}_i^{(k,\bar{k})} \right\vert \, ,
\label{distance}
\ee
where $\bar{x}_i^{(k,\bar{k})}$ is the (interpolated) $x$-coordinate of a point in curve $\bar{k}$ with $y$-coordinate equal to $y_i^{(k)}$, i.e. the projection of point $(x_i^{(k)},y_i^{(k)})$ of curve $k$ on curve $\bar{k}$ along the $x$-axis. The innermost sum over $i$ in Eq. (\ref{distance}) is restricted to points in curve $k$ which overlap with curve $\bar{k}$ along the $y$-direction, i.e. those points in $k$ whose $y$-coordinate is between the minimum and maximum $y$-coordinate of curve $\bar{k}$. In order to obtain now the projection $\bar{x}_i^{(k,\bar{k})}$ in Eq. (\ref{distance}) any interpolation scheme can be used, though for our purposes the simplest linear interpolation works well. In particular, we choose
\be
\bar{x}_i^{(k,\bar{k})} = \frac{y_i^{(k)} - B_i^{(k,\bar{k})}}{A_i^{(k,\bar{k})}} \, ,
\label{proj}
\ee
with $A_i^{(k,\bar{k})}$ and $B_i^{(k,\bar{k})}$ the slope and the $y$-intercept of the interpolating function,
\ben
A_i^{(k,\bar{k})} &=& \frac{ y_{i+}^{(\bar{k})} - y_{i-}^{(\bar{k})} }{ x_{i+}^{(\bar{k})} - x_{i-}^{(\bar{k})} } \, , \nonumber \\
B_i^{(k,\bar{k})} &=& \frac{ y_{i+}^{(\bar{k})} x_{i-}^{(\bar{k})} - y_{i-}^{(\bar{k})}x_{i+}^{(\bar{k})}}{ x_{i+}^{(\bar{k})} - x_{i-}^{(\bar{k})} } \, . \nonumber
\een
The points $(x_{i\pm}^{(\bar{k})},y_{i\pm}^{(\bar{k})})$ correspond to the points in the $\bar{k}$-curve bracketing point $i$ of $k$-curve along the $y$-direction, see sketch in Fig. \ref{fig11}. To normalize the distance metric, we divide the resulting sums by the total number of overlapping points, ${\cal N}_{\text{overl}}$. Moreover, because the $L^{-\alpha}$ scaling in the $x$-coordinate of the measured density and temperature profiles may affect strongly the total span of the collapsed curves depending on the anomaly exponent $\alpha$ used, the collapse metric is also normalized by the total span in the $x$-direction of the curve cloud, $\ell_{\text{max}}\equiv (x_{\text{max}}-x_{\text{min}})$ with $x_{\text{max}}=\max_{k,i} [\{x_i^{(k)}\},i\in[1,M], k\in[1,K]]$ and $x_{\text{min}}=\min_{k,i} [\{x_i^{(k)}\},i\in[1,M], k\in[1,K]]$, i.e. our distance is relative to the total span of the curve cloud in the $x$-direction.

In order to obtain the exponent $\alpha$ characterizing anomalous Fourier's law in our $1d$ fluid, we minimize the metric (\ref{distance}) for varying mass ratios $\mu$. In fact, the collapse metric $D(\alpha,\mu)$ exhibits a deep and narrow minimum as a function of $\alpha$ for each $\mu$, see inset to Fig. 3 in the main text, offering a precise measurement of the anomaly exponent. Moreover, an estimate of the exponent error can be obtained from the width and depth of this minimum \cite{bhattacharjee01aA}. By expanding $\ln D(\alpha,\mu)$ around the minimum at $\alpha=\alpha_0$, the width can be estimated as \cite{bhattacharjee01aA}
\be
\Delta \alpha = \frac{\displaystyle \epsilon \alpha_0}{\displaystyle \sqrt{2 \ln \left[ \frac{D(\alpha_0 \pm \epsilon \alpha_0,\mu)}{D(\alpha_0 ,\mu)}\right]}} \, ,
\label{error}
\ee
for a given level $\epsilon$. Here we choose $\epsilon=0.01$, so the estimate for the anomaly exponent is $\alpha_0\pm \Delta\alpha$ with an errorbar reflecting the width of the minimum at the $1\%$ level \cite{bhattacharjee01aA}.


\end{document}